\documentclass[lettersize,journal]{IEEEtran}
\usepackage{amsmath,amsfonts}
\usepackage{algorithmic}
\usepackage{algorithm}
\usepackage{array}
\usepackage[caption=false,font=normalsize,labelfont=sf,textfont=sf]{subfig}
\usepackage{textcomp}
\usepackage{stfloats}
\usepackage{url}
\usepackage{verbatim}
\usepackage{graphicx}
\usepackage{cite}
\usepackage{svg}
\usepackage{xcolor}
\usepackage[printonlyused]{acronym}
\usepackage{caption}

\newcommand{\m}[1]{\mathbf{#1}}

\newcommand{\hermitian}[0]{\text{H}}
\newcommand{\transpose}[0]{\text{T}}
\newcommand{\bluetext}[1]{{\color{black}#1}}

\newtheorem{theorem}{\textbf{Theorem}}

\newtheorem{lemma}{\textbf{Lemma}}

\newtheorem{remark}{\textbf{Remark}}

\acrodef{ANM}{atomic norm minimization}
\acrodef{DoA}{direction of arrival}
\acrodef{AST}{atomic norm soft thresholding}
\acrodef{KKT}{Karush–Kuhn–Tucker}
\acrodef{ADMM}{alternating direction method of multiplier}
\acrodef{SDP}{semi-definite programming}
\acrodef{SVD}{singular-value decomposition}
\acrodef{FFT}{Fast Fourier Transform}
\acrodef{BFGS}{Broyden–Fletcher–Goldfarb–Shanno}
\acrodef{DFT}{Discrete Fourier Transform}
\acrodef{LASSO}{Least Absolute Shrinkage and Selection Operator}
\acrodef{SOTA}{state of the art}
\acrodef{NOMP}{Newtonized orthogonal matching pursuit}
\acrodef{MMV}{multiple measurement vector}
\acrodef{CD}{coordinate descent}
\acrodef{CRB}{Cramér–Rao bound}
\acrodef{SNR}{signal-to-noise ratio}
\acrodef{OFDM}{orthogonal frequency division multiplexing}

\begin{document}

\title{A Coordinate Descent Approach to Atomic Norm \bluetext{Denoising}}

\author{Ruifu~Li, ~Danijela~Cabric,~\IEEEmembership{Fellow,~IEEE}%
\thanks{Ruifu Li and Danijela Cabric are with the Electrical and Computer Engineering Department, University of California, Los Angeles, Los Angeles, CA 90095 (e-mail: doanr37@ucla.edu; danijela@ee.ucla.edu).}
\thanks{This work is supported by NSF under grant 1718742 and 1955672.}}


\maketitle

\thispagestyle{empty}
\pagestyle{empty}

\begin{abstract}
Atomic norm minimization is of great interest in various applications of 
sparse signal processing including super-resolution line-spectral estimation and 
signal denoising. In practice, \ac{ANM} is formulated as \ac{SDP} that is generally hard to solve. 
This work introduces a low-complexity {\color{black}solver for a type of \ac{ANM} known as \ac{AST}}. 
The proposed method uses the framework of 
coordinate descent and exploits the sparsity-inducing nature of atomic-norm regularization. Specifically, 
this work first provides an equivalent, non-convex formulation of \ac{AST}. It is then proved that applying a coordinate 
descent algorithm on the non-convex formulation {\color{black} leads to convergence to the
global solution}. For the case of   
a single measurement vector of length $N$ and {\color{black} complex exponential basis}, the complexity of each step in the coordinate descent procedure is 
$\mathcal{O}(N\log N)$, rendering the method efficient for large-scale problems. {\color{black} Through simulations, the proposed solver 
is shown to be faster than \ac{ADMM} or customized interior point \ac{SDP} solver if the problems are sparse}. It is demonstrated that 
the coordinate descent {\color{black} solver} 
can be modified for {\color{black}\ac{AST} with
 multiple dimensions and multiple measurement vectors as well as a variety of general basis}. 
\end{abstract}


\section{Introduction}
\IEEEPARstart{I}{n} the post compressed sensing era, \ac{ANM} is a powerful candidate for finding a sparse representation of 
a measured signal, as it resolves the basis mismatch problem \cite{chi2011sensitivity} that typically arises for \ac{DFT} basis. 
The advantage of atomic norm originates from {\color{black} its connection with a continuous manifold typically known as the 
atomic set}. As opposed to a basis of a finite number of vectors in conventional $\ell_1$-norm regularized least-square 
regression, the atomic set contains infinitely many vectors. Consequently, the sparse reconstruction from an atomic norm 
regularized least-squares regression can consist of any points on a continuous manifold. This feature makes \ac{ANM} a 
powerful tool on estimations of continuous parameters (delay, frequencies, Doppler, etc.), as well as denoising of 
signals such as images or speeches.

The price to pay {\color{black} for searching over a continuous dictionary} is the amount of computation required to {\color{black}reach a solution}. 
Unlike constrained least-squares problems, \ac{ANM} is originally formulated 
as a \ac{SDP} in the seminal paper \cite{candes2014towards} based on the bounded real lemma \cite{magazine} that 
characterizes the infinite-dimensional constraints. From then, the \ac{SDP} formulation of \ac{ANM} has been applied 
extensively to various estimation problems. For a $N$-dimensional complex vector $\m{y}\in\mathbb{C}^N$, the 
cost of solving \ac{ANM} is $O(N^4)$ if a general purpose interior-point solver is used \cite{cvxpy}, and $O(N^3)$ with 
the customized interior-point solver \cite{hansen2019fast} or with the proximal methods \cite{wang2018ivdst}. For the 
\ac{ANM} problem of \ac{MMV} $\m{y}_1, \m{y}_2, ..., \m{y}_M$, the cost increases to $\mathcal{O}(N + M)^3$ \cite{exactjoint}. 
Such {\color{black}high computational complexity posts serious limitations on applicability of \ac{ANM} for large scale problems} 
where $N$ or $M$ is on the order of $10^4$ or higher. In spite of this, \ac{ANM} is still considered to be one of 
most popular tool in {\color{black} applications of super-resolution}. Over the years numerous variants of \ac{ANM} are developed, 
including \ac{ANM} with decoupled formulations \cite{decoupled2021}, with multidimensional frequency estimation \cite{vandermonde}, and with weighted atomic set \cite{reweighted2016}. 
It has also been applied to signal denoising \cite{ast}, linear system identification \cite{identification2012}, and wireless channel estimation \cite{channel2019}, etc.  

To fill the gap on efficient solvers of \ac{ANM}, this work provides an iterative, low-complexity framework for solving {\color{black} \ac{AST}, a specific form of \ac{ANM} 
that can be interpreted as an atomic norm regularized least-squares regression}. The {\color{black}solver} is based on coordinate descent algorithm. 
It {\color{black}utilizes} a mixed-integer but equivalent formulation of the \ac{AST} which shares the same global optimal point with the corresponding \ac{SDP} formulation. 
{\color{black}Additionally, this work provides a simple proof that the coordinate descent solver would 
asymptotically converge to the global solution of {\color{black}\ac{AST}}}. 
The main advantages of the proposed {\color{black}solver} include: 
\begin{itemize}
    \item For the classic basis of \ac{DFT} vectors, the {\color{black} solver has low complexity per iteration}. With \ac{FFT}, the cost 
    per-iteration is $\mathcal{O}(N\log N)$.
    \item The {\color{black} solver applies to} a variety of \ac{AST} problems, {\color{black} including those with multiple dimensions and multiple measurements}. 
    Theoretically, \bluetext{the solver can adapt to all atomic sets $\mathcal{A}$ for  which projections onto their conic sets $\left\{c\m{a}\left| \m{a} \in \mathcal{A}, c\in \mathbb{C}\right.\right\}$ can be evaluated}.
    \item The {\color{black} solver is simple to implement as it does not rely on \ac{SDP}}. 
    \item The {\color{black} solver} is {\color{black} empirically observed to have rapid convergence when solutions are sparse}.
\end{itemize}

 
The rest of the paper is organized as follows. In section II, we briefly introduce preliminary background and related work. 
{\color{black} The theoretical foundation behind the proposed coordinate descent solver is discussed in section III, while the algorithmic design as well as its 
convergence based on such foundations are discussed in section IV.
The extension of the method to various \ac{AST} problems are presented in section V. Section VI presents a short discussion and suggested future works. The paper is concluded in section VII.}

\textit{Notations:} Scalars, vectors, and matrices are denoted by non-bold,
 bold lower-case, and bold upper-case letters, respectively, e.g. $h$, $\mathbf{h}$ 
 and $\mathbf{H}$. The element in $i$-th row and $j$-th column in matrix $\mathbf{H}$ is 
 denoted by $[\mathbf{H}]_{i,j}$. Transpose and Hermitian transpose are denoted by $(.)^{\transpose}$ 
 and $(.)^{\hermitian}$, respectively. The $l_p$-norm of a vector $\mathbf{h}$ is denoted 
 by \bluetext{$||\mathbf{h}||_{\ell_p}$.} The symbol $\text{diag}(\mathbf{A})$ aligns diagonal elements of $\mathbf{A}$ into a vector, 
 and $\text{diag}(\mathbf{a})$ aligns vector $\mathbf{a}$ into a diagonal matrix. The operator $\mathcal{T}\left(.\right)$ 
 denotes the mapping from a vector to a Toeplitz matrix whose first column being the vector provided. The inner product 
 between two elements $\m{x}, \m{y}$ from a vector space $\m{x}, \m{y}$ is denoted as $\left<\m{x}, \m{y}\right>$. 
Unless otherwise stated, it is assumed that the $\ell_2$-norm can be induced by the inner product, i.e., 
$\left\|\m{x}\right\|_2^2 = \left<\m{x}, \m{x}\right>$. For a countable set $\mathcal{S}$, $|\mathcal{S}|$ denote 
the number of elements in $\mathcal{S}$, while $\left[\mathcal{S}\right]_i$ denote the $i$-th element from the set.

\section{Preliminaries and Related Work}

In this section the basic concepts of atomic set and atomic norm are introduced. A brief overview on the applications of \ac{ANM} and related work follows. 

In short, the atomic norm generalizes the notion of $\ell_1$-norm to a continuous basis. In an $N$-dimensional vector space,  
consider the classical problem of representing a signal $\m{y}$ with elements $\m{a}_i$ from a set of basis $\mathcal{A}$. With the celebrated \ac{LASSO} regression, the basis set $\mathcal{A}$ is usually 
finite and over complete which can be represented as a matrix $\m{A} = \left[\m{a}_0, \m{a}_1, ..., \m{a}_i, ... \right]$. A sparse representation can be induced simply by adding $\ell_1$-norm regularization, leading to the following optimization problem: 
\begin{equation}
\begin{split}
  \underset{\m{c}}{\mathrm{Minimize}} & \quad \left\|\m{c}\right\|_1 + \frac{\zeta}{2}\left\|\m{y} - \m{A}\m{c}\right\|_2^2\\
\end{split}
\label{eq:lasso}
\end{equation}
where $\zeta > 0$ is a positive constant that balances the error and the sparsity of the solution. 
When the basis set $\mathcal{A}$ contains infinitely many elements, a matrix presentation like (\ref{eq:lasso}) is no 
longer available. Instead of $\ell_1$-norm, the regularization is now based on the 
atomic norm $\left\|.\right\|_\mathcal{A}$. The correspondence of (\ref{eq:lasso}) with an infinite-dimensional set
 of basis $\mathcal{A}$ is the \ac{AST} \cite{ast}:  
\begin{equation}
\begin{split}
  \underset{\m{x}}{\mathrm{Minimize}} & \quad \left\|\m{x}\right\|_\mathcal{A} + \frac{\zeta}{2}\left\|\m{y} - \m{x}\right\|_2^2\\
\end{split}
\label{eq:anm}
\end{equation}
with $\left\|\m{x}\right\|_\mathcal{A}$ defined as: 
\begin{equation}\label{eq:atomic_norm_v1}
    \left\|\m{x}\right\|_{\mathcal{A}} = \inf \left\{t > 0\left|\ \m{x} \in t \cdot \mathrm{conv}(\mathcal{A})\right.\right\}
\end{equation}
The definition (\ref{eq:atomic_norm_v1}) is abstract\footnote{\cite{magazine} delivers an excellent and friendly exposition to \ac{ANM} that includes more detailed and rigorous discussions on the concept of atomic norm.} as it involves the concept of a convex hull $\mathrm{conv}(\mathcal{A}) $ of
the infinite-dimensional set $\mathcal{A}$. Fortunately, as long as the set $\mathcal{A}$ is symmetrical 
with respect to the origin, the definition 
(\ref{eq:atomic_norm_v1}) can be equivalently interpreted from a total variation perspective \cite{magazine}: 
\begin{equation}\label{eq:atomic_norm_v2}
    \left\|\m{x}\right\|_{\mathcal{A}} = \inf \left\{\sum\limits_i c_i\left|\m{x} = \sum\limits_ic_i\m{a}_i,\ c_i > 0,\ \m{a}_i \in \mathcal{A}\right.\right\}
\end{equation}
with the definition (\ref{eq:atomic_norm_v2}), a mixed-integer formulation of \ac{AST} can be derived accordingly: 
\begin{equation}\label{eq:anm_v2}
\begin{split}
  \underset{{\color{black}\begin{smallmatrix}
    L,\\ 
    c_i \geq 0,\\
    \m{a}_i \in \mathcal{A}
  \end{smallmatrix}}}{\mathrm{Minimize}} & \quad \sum\limits_{i}^L c_i + \frac{\zeta}{2}\left\|\m{y} - \sum\limits_{i}^L c_i\m{a}_i\right\|_2^2\\
\end{split}
\end{equation}
{\color{black} The mixed integer representation is intuitively a representation of the original \ac{AST} (\ref{eq:anm}) in the parameter space of the atomic set}. In section \ref{sec:algorithm}, the two problems, {\color{black}(\ref{eq:anm}) and (\ref{eq:anm_v2})}, are proved to be equivalent as they shared the same {\color{black} global minimal objective}. 

The notion of \ac{AST} can be better motivated with its applications. Some most frequently used atomic sets include:
\begin{itemize}
    \item{The set of complex exponential vectors. The set consists of continuous \ac{DFT} basis defined with a frequency $f$ over $[0, 2\pi)$: 
    \begin{equation}\label{eq:atomic_set_dft}
      \left\{e^{\mathrm{1j}\phi}\m{a}\left(f\right)\left|\phi, f \in \left[0, 2\pi\right), \left[\m{a}\left(f\right)\right]_i = e^{\mathrm{1j}(i - 1)f}\right.\right\}
    \end{equation} 
    The set naturally arises in a variety of array signal processing and wireless communication problems.
    }
    \item{\ac{DFT} basis with multiple snapshots. The set consists of continuous \ac{DFT} basis combined with a unit-norm sphere via the outer-product: 
    \begin{equation}\label{eq:atomic_set_dft_mmv}
      \left\{\m{a}\left(f\right) \otimes \m{c} \left| \left[\m{a}\left(f\right)\right]_i = e^{\mathrm{1j}(i - 1)f}, \left\|\m{c}\right\|_2 = 1\right.\right\}
    \end{equation} 
    The set plays a major role in estimation problems with \ac{MMV} where the observation $\m{y}$ in (\ref{eq:anm}) {\color{black} becomes a matrix $\m{Y}$ instead of a vector}.
    }
    \item{Two-dimensional \ac{DFT} basis. The set consists of 2D continuous \ac{DFT} basis that naturally arises in signal processing problems with a uniform planar antenna array 
    \begin{equation}\label{eq:atomic_set_dft_2d}
      \left\{e^{\mathrm{1j}\phi}\m{a}\left(f_1\right) \otimes \m{a}\left(f_2\right)\left| \left[\m{a}\left(f\right)\right]_i = e^{\mathrm{1j}(i - 1)f}\right.\right\}
    \end{equation} 
    }
\end{itemize}

Plugging either of these atomic sets (\ref{eq:atomic_set_dft}) - (\ref{eq:atomic_set_dft_2d}) into the \ac{AST} (\ref{eq:anm}) produces a well-defined convex optimization problem 
with semi-definite constraints. For instance, with $\mathcal{A}$ defined as (\ref{eq:atomic_set_dft}), (\ref{eq:anm}) is equivalent to the following \ac{SDP} \cite{chi2014joint}: 
\begin{equation}\label{eq:anm_sdp}
\begin{split}
 \underset{\m{x}, \m{u}, t}{\mathrm{Minimize}} & \quad \frac{t}{2} + \frac{1}{2N}\mathrm{tr}\left(\mathcal{T}(\m{u})\right) + \frac{\zeta}{2}\left\|\m{y} - \m{x}\right\|_2^2\\
  \mathrm{subject\ to} & \quad  \left[\begin{matrix}
    \mathcal{T}\left(\m{u}\right) & \m{x} \\
    \m{x}^\hermitian & t \\
  \end{matrix}\right] \succeq 0\\
\end{split}
\end{equation}
The conversion between $(\ref{eq:anm})$ to $(\ref{eq:anm_sdp})$ is non-trivial as it relies on the fact that the outer product 
of a \ac{DFT} vector with itself produces a hermitian Toeplitz matrix, i.e., 
$\mathcal{T}\left(\m{a}(f)\right) = \m{a}(f)\m{a}^\hermitian\left(f\right)$. Therefore, 
the SDP formulation $(\ref{eq:anm_sdp})$ cannot be generalized to arbitrary atomic sets. Nevertheless, the 
legitimate \ac{SDP} (\ref{eq:anm_sdp}) can be solved in polynomial time with several existing algorithms. 
\cite{magazine} proposed the well-known \ac{ADMM} method to tackle the semi-definite constraints. In each iteration, 
the complexity of the projection step which requires a \ac{SVD} is $\mathcal{O}\left(N^3\right)$. 
On the other hand, \cite{hansen2019fast} addressed a customized primal-dual interior point 
solver for (\ref{eq:anm_sdp}). The computation of the exact Hessian matrix inevitably requires $\mathcal{O}\left(N^3\right)$ {\color{black} operations}. 
\cite{wang2018ivdst} proposed the proximal gradient method for a variant of (\ref{eq:anm_sdp}) with 
hard thresholding. The method has $\mathcal{O}(N^2)$ complexity per iteration. Different from the works above, rather than focusing 
on the \ac{SDP} (\ref{eq:anm_sdp}), this work addresses the mixed-integer formulation (\ref{eq:anm_v2}) directly. 
From such a perspective, the most related method from previous literature is the \ac{NOMP} \cite{newtonized}. 
The major limitation of \ac{NOMP} is that instead of solving the \ac{AST}, it {\color{black} is designed to solve problems with atomic 
$\ell_0$ norm $\left\|.\right\|_{\mathcal{A},0}$ regularization}. Consequently, due to the highly non-convex 
nature of the $\ell_0$ norm, \ac{NOMP} is not guaranteed to converge to the global optimal point. The discussion 
in \cite{newtonized} is also limited to the continuous \ac{DFT} basis (\ref{eq:atomic_set_dft}), (\ref{eq:atomic_set_dft_mmv}) 
and doesn't extend to general atomic sets.

{\color{black} In conventional \ac{LASSO} regression, the application of coordinate descent method has been} discussed 
by \cite{friedman2010regularization}, \cite{tibshirani2011solution}. Although most discussions are 
limited to the case of a fixed basis (\ref{eq:lasso}), they provide inspiration to our algorithmic development in 
Section \ref{sec:algorithm}. In the next few sections, the equivalence between {\color{black} the two} different \ac{AST} formulations {\color{black} (\ref{eq:anm_v2}) and (\ref{eq:anm})} is 
established, {\color{black}which provides the theoretical foundation for the design of the proposed solver}.
    
{\color{black}\section{Theoretical Equivalence Between the Two \ac{AST} Formulations}}

{\color{black} This section establishes the theoretical foundation behind the proposed coordinate descent solver. 
By proving that the two formulations of \ac{AST} are equivalent, the sufficient and necessary condition for reaching global 
solution of (\ref{eq:anm}) is then naturally translated to the corresponding condition of (\ref{eq:anm_v2}). 
Specifically, the current section takes three steps in establishing the equivalence and translating the conditions. As a start, Lemma \ref{lemma: optimality} restates condition \cite[Lemma 1]{ast} of reaching optimal point for the original \ac{AST} formulation. 
Then, Lemma \ref{lemma:single_atom} and Theorem \ref{theorem:anm_equivalence} provide the equivalence between the two formulations 
(\ref{eq:anm}) and (\ref{eq:anm_v2}). Finally, Theorem \ref{theorem:anm_v2_optimality} establishes the condition for reaching global solution of (\ref{eq:anm_v2}).}

{\color{black} The following condition is both sufficient and necessary for the solution of original \ac{AST} (\ref{eq:anm})} \cite[Lemma 1]{ast}:
\begin{lemma}\label{lemma: optimality}
{\color{black}$\m{x}^*$ is the solution to the optimization problem (\ref{eq:anm}) if and
 only if: (i) $\sup_{\m{a}\in\mathcal{A}}\left<\m{y} - \m{x}^*, \m{a}\right> \leq \zeta^{-1} $  (ii) $\left<\m{x}^*, \m{y} - \m{x}^*\right> = \zeta^{-1} \left\|\m{x}^*\right\|_\mathcal{A}$.}
\end{lemma}

Let $\m{z}^* = \m{y} - \m{x}^*$ be the residual of $\m{y}$ in the solution. Then $\m{z}^*$ is also known as the dual certificate of support as it indicates {\color{black}the presence of} elements $\m{a}_i$ in the sparse representation of $\m{x}^*$. In Lemma \ref{lemma: optimality}, if the solution $\m{x}^*$ is non-trivial, i.e., $\m{x}^*\neq \m{0}$, then the inequality in (i) is tight. Consequently, there exist elements $\m{a}_i \in \mathcal{A}$ such that $\left<\m{y} - \m{x}^*, \m{a}_i\right> = \zeta^{-1}$. Let $\mathcal{S}$ be the set of all such elements. The solution  $\m{x}^*$ then admits  a decomposition over $\mathcal{S}$: $\m{x}^* = \sum\limits_i c_i \m{a}_i, \m{a}_i \in \mathcal{S}$ \cite{candes2014towards}. 
Such a decomposition of $\m{x}^*$ is unique 
due to the existence of the dual certificate $\m{z}^*$ 
\cite[Corollary 1]{ast}. In a typical 
use case of \ac{AST} such as \ac{DoA} 
estimation \cite{raj2018single}, the elements $\m{a}_i \in \mathcal{S}$ often reveal the values of 
the estimated parameter from its continuous domain.    

With the notion of the dual {\color{black} certificate}, the equivalence between (\ref{eq:anm}) and 
(\ref{eq:anm_v2}) can be readily shown. {\color{black}In the original \ac{AST} (\ref{eq:anm}), the residual $\m{z}^* = \m{y} - \m{x}^*$ is the dual certificate. Then, in the mixed integer formulation, 
a natural guess is that $\m{y} - \sum_{i}^L c_i\m{a}_i$ is {\color{black} effectively the dual 
certificate of support}. To see this, the following Lemma 2 states a common property shared by $\m{y}_r$ and $\m{z}^*$:} 
\begin{lemma}\label{lemma:single_atom}
    Suppose $\mathcal{A}$ is symmetric with respect to the origin. Let $(c_1, \m{a}_1), (c_2, \m{a}_2),..., (c_L, \m{a}_L)$ be the global optimal point to the mixed integer problem (\ref{eq:anm_v2}) such that $c_i > 0, \m{a}_i \in \mathcal{A}$. Then the residual $\m{y}_r = \m{y} - \sum _{i}^L c_i\m{a}_i$ must satisfy the inequality: 
    $ \sup_{a\in\mathcal{A}}\left<\m{y}_r, \m{a}\right> \leq \zeta^{-1}$.
\end{lemma}

\begin{IEEEproof}
    The key is to consider a function $f\left(\m{x}, \mathcal{A}, \zeta\right)$ defined over the vector space: 
    \begin{align} \label{eq:single_atom}
        f\left(\m{x},\mathcal{A}, \zeta\right) & = \inf_{c \geq 0, \m{a}\in\mathcal{A}} 
        \frac{\zeta}{2}\left\|\m{x} - c\m{a}\right\|_2^2 + c  \\ \nonumber
            & = \frac{\zeta}{2} \left\|\m{x}\right\|_2^2 + \inf_{c \geq 0, \m{a}\in\mathcal{A}} c\left(1 - \zeta\left<\m{x}, \m{a}\right>\right) + \frac{c^2\zeta}{2}\left\|\m{a}\right\|_2^2 \\ \nonumber 
            & \leq \frac{\zeta}{2} \left\|\m{x}\right\|_2^2 
    \end{align}
    Notice that (\ref{eq:single_atom}) is exactly a shrinkage and thresholding operation. When the inequality 
    $\sup_{a\in\mathcal{A}}\left<\m{x}, \m{a}\right> \leq \zeta^{-1}$ holds, 
    $ f\left(\m{x},\mathcal{A}, \zeta\right) =  \frac{\zeta}{2} \left\|\m{x}\right\|_2^2 $. And the reverse statement is also true. If $ f\left(\m{x},\mathcal{A}, \zeta\right) =  \frac{\zeta}{2} \left\|\m{x}\right\|_2^2 $, there must be $\sup_{a\in\mathcal{A}}\left<\m{x}, \m{a}\right> \leq \zeta^{-1}$.
    
    Since the set of tuples $(c_i, \m{a}_i), i = 1,..., L$ is the global optimal point, its objective value must be the global minimum. Therefore, adding one more tuple $(c_0, \m{a}_0)$ to the set can only increase the objective value. This results in the following inequality: 
    \begin{align}\label{eq:optimal_L}\nonumber
        \sum\limits_{i = 1}^L c_i + \frac{\zeta}{2}\left\|\m{y}_r\right\|_2^2 & \leq \inf_{c_0 \geq 0, \m{a}_0\in\mathcal{A}} \sum\limits_{i = 0}^{L} c_i 
        +   
        \frac{\zeta}{2}\left\|\m{y}_r - c_0\m{a}_0\right\|_2^2 \\ \nonumber 
        & = \sum\limits_{i = 1}^L c_i + f\left(\m{y}_r,\mathcal{A}, \zeta\right) \\ \nonumber
        & \leq \sum\limits_{i = 1}^L c_i + \frac{\zeta}{2}\left\|\m{y}_r\right\|_2^2
    \end{align}
    It remains trivial to see that $f\left(\m{y}_r,\mathcal{A}, \zeta\right) = \frac{\zeta}{2}\left\|\m{y}_r\right\|_2^2$, which means $\sup_{a\in\mathcal{A}}\left<\m{y}_r, \m{a}\right> \leq \zeta^{-1}$.
\end{IEEEproof}

Lemma \ref{lemma:single_atom} points out that in the solution of (\ref{eq:anm_v2}) 
the residual $\m{y}_r$ must satisfy the condition (i) in Lemma \ref{lemma: optimality}. It remains to show that condition (ii) should also be satisfied. Condition (ii) relies on the fact that each tuple $(c_i, \m{a}_i)$ in the solution must also be a stationary point. 
Let $h$ be the objective function in (\ref{eq:anm_v2}), i.e., 
\begin{equation}\label{eq:anm_v2_obj}
    h(c_1, \m{a}_1, .., c_L, \m{a}_L) = \sum\limits_{i = 1}^L c_i + \frac{\zeta}{2}\left\|\m{y} - \sum\limits_{i}^L c_i\m{a}_i\right\|_2^2 
\end{equation}
Since each tuple is a stationary point of $h$, their partial derivative must be $0$. Specifically, let $\m{y}_r^{i} = \m{y}_r + c_i\m{a}_i$. The following condition on partial derivative must be satisfied: 
\begin{equation}\label{eq:anm_v2_deri_c}
    \begin{split}
        \frac{\partial{h}}{\partial c_i} & = \frac{\partial}{\partial c_i}\left(\frac{\zeta}{2}\left\|\m{y}_r^i - c_i\m{a}_i\right\|_2^2 + \sum\limits_{j = 1}^L c_j \right) \\
        & = -\zeta\left<\m{a}_i, \m{y}_r^i - c_i\m{a}_i\right> + 1\\
        & = 0 \\
    \end{split}
\end{equation}
Which implies that: 
\begin{equation}\label{eq:anm_condition_c}
    \begin{split}
       \left<\m{a}_i,\m{y}_r^i\right> - \zeta^{-1} = c_i \left\|\m{a}_i\right\|_2^2
    \end{split}
\end{equation}
Notice that it is defined such that: $\m{y}_r^i = \m{y}_r + c_i\m{a}_i$. Therefore, by plugging in the definition of $\m{y}_r$ into (\ref{eq:anm_condition_c}), 
\begin{equation}\label{eq:anm_condition_a}
    \begin{split}
       \left<\m{a}_i,\m{y}_r\right> = \zeta^{-1} 
    \end{split}
\end{equation}
\begin{remark}
    (\ref{eq:anm_condition_a}) shows that {\color{black} $\m{y}_r$ has the indicating property of dual certificate, which means the inequality in Lemma \ref{lemma:single_atom} is tight. $\m{y}_r$ has the maximum inner product $\left<\m{y}_r, \m{a}\right> = \zeta^{-1}$ with $\m{a} \in \mathcal{A}$ if $\m{a}$  is part of the solution.} Using this property, the equivalence between (\ref{eq:anm}) and (\ref{eq:anm_v2}) can be readily established.  
\end{remark}
\begin{theorem}\label{theorem:anm_equivalence}
     Suppose $\mathcal{A}$ is symmetric with respect to the origin. Let $(c_1, \m{a}_1), (c_2, \m{a}_2),..., (c_L, \m{a}_L)$ be {\color{black}the global solution} to the mixed integer problem (\ref{eq:anm_v2}) such that $c_i > 0, \m{a}_i \in \mathcal{A}$. Then $\m{x} = \sum_i^L c_i\m{a}_i$ is also the solution to the {\color{black}\ac{AST}} problem 
     (\ref{eq:anm}) and $\left\|\m{x}\right\|_\mathcal{A} = \sum_i^L c_i$. 
\end{theorem}

\begin{IEEEproof}
    With Lemma \ref{lemma:single_atom}, the first condition in \ref{lemma: optimality} has been proved to be satisfied by the set of tuples $(c_i, \m{a}_i)$. To show that the second condition is also satisfied by $\m{x} = \sum_i^L c_i\m{a}_i$, the first step is to use the property (\ref{eq:anm_condition_a}). Notice that: 
    \begin{align}\label{anm_condition_norm}
        \left<\m{y} - \m{x}, \m{x}\right>  
        = \sum\limits_{i}^L c_i \left<\m{y}_r, \m{a}_i\right> = \frac{1}{\zeta}\sum\limits_{i = 1}^L 
        c_i  
    \end{align}
    It remains to show that $\sum_{i}^L c_i = \left\|\m{x}\right\|_\mathcal{A}$. Since the set of tuples is the global optimal point that solves (\ref{eq:anm_v2}), it must satisfy the definition (\ref{eq:atomic_norm_v2}), i.e., $\sum_{i}^L c_i = \left\|\m{x}\right\|_\mathcal{A}$ must hold. Consequently, $\m{x} = \sum_{i}^L c_i\m{a}_i$ satisfies both the conditions of optimality for (\ref{eq:anm}) 
    in Lemma \ref{lemma: optimality}. This concludes the proof.
\end{IEEEproof}

\begin{remark}\label{remark:anm_v2_condition}
{\color{black} A direct consequence of Theorem \ref{theorem:anm_equivalence} is that the optimal objective value 
of (\ref{eq:anm_v2}) is the same as that of (\ref{eq:anm}). This is because the solution of 
(\ref{eq:anm_v2}) solves (\ref{eq:anm}), and the difference between the objective in (\ref{eq:anm}) and in (\ref{eq:anm_v2}) 
is only the difference 
between  $\sum_{i}^L c_i$ and $\left\|\m{x}\right\|_\mathcal{A}$, which is $0$ for 
every solution of (\ref{eq:anm_v2}). 
This fact is useful when establishing the condition for reaching a 
global solution of (\ref{eq:anm_v2}) as stated formally in the following theorem:} 
\end{remark}

{\color{black}
\begin{theorem}\label{theorem:anm_v2_optimality}
    Suppose $\mathcal{A}$ is symmetric with respect to the origin. 
    A set of tuples $\mathcal{S} = \{(c_1, \m{a}_1), (c_2, \m{a}_2),..., (c_L, \m{a}_L)\}$ 
    is {\color{black}the global solution} to the mixed 
    integer problem (\ref{eq:anm_v2}) if and only if  
    (i) $\sup_{\m{a}\in\mathcal{A}}\left<\m{y} - \sum_{i = 1}^L c_i\m{a}_i, \m{a}\right> 
    \leq \zeta^{-1} $  
    (ii) $\left<\sum_{i = 1}^L c_i\m{a}_i, \m{y} - \sum_{i = 1}^L c_i\m{a}_i\right> = \zeta^{-1} 
    \sum_{i = 1}^L c_i$.    
\end{theorem}
\begin{IEEEproof}
   Let $\m{y}_r = \m{y} - \sum_{i = 1}^L c_i\m{a}_i$. Lemma \ref{lemma:single_atom} establishes the 
   first condition in the forward statement, i.e., given $\mathcal{S}$ being optimal, 
   $\sup_{\m{a}\in\mathcal{A}}\left<\m{y}_r, \m{a}\right> 
   \leq \zeta^{-1} $. The second condition in the forward direction is established by 
   (\ref{eq:anm_v2_deri_c}) - (\ref{eq:anm_condition_a}).
   
   To establish the statement in the backward direction, the first step is to show that any set 
   $\mathcal{S}$ that satisfies (i) and (ii) would solve the original \ac{AST} problem. 
   In this step, the key is to  
   establish: 
   \begin{equation}\label{eq:dual_support}
    \left\|\sum_{i = 1}^L c_i\m{a}_i\right\|_\mathcal{A} = \sum_{i = 1}^L c_i 
   \end{equation}   
   This is in fact a 
   classic result in the framework of super-resolution \cite{candes2014towards}, \cite{exactjoint}. 
   For readers' convenience, an outline of the proof is provided. By definition,
   $\left\|\sum_{i = 1}^L c_i\m{a}_i\right\|_\mathcal{A} \leq \sum_{i = 1}^L c_i$.
   Now suppose $\left\|\sum_{i = 1}^L c_i\m{a}_i\right\|_\mathcal{A} < \sum_{i = 1}^L c_i$.  
   Consequently, $\sum_{i = 1}^L c_i\m{a}_i$ admits a different decomposition
   $\sum_{i = 1}^L c_i\m{a}_i = \sum_{i}^{L'}c_i'\m{a}_i'$
   such that $\sum_i^{L'} c_i' = \left\|\m{x}\right\|_\mathcal{A} < \sum_i^L c_i$. 
   An inequality is then established using (i) and (ii):
    \begin{equation}\label{eq:anm_conflict}
    \begin{split}
        \left<\m{y}_r, \sum\limits_i^{L'} c_i'\m{a}_i'\right> 
            \leq \zeta^{-1}\sum\limits_i^{L'} c_i' < \zeta^{-1}\sum\limits_i^{L} c_i
            = \left<\m{y}_r, \sum_{i = 1}^L c_i\m{a}_i\right>
        \end{split}
    \end{equation}
    which results in a conflict since the leftmost quantity is the same as the rightmost quantity.
    Therefore, (\ref{eq:dual_support}) holds. 
    
        The three equations  
    (\ref{eq:dual_support}), (i) in Theorem \ref{theorem:anm_v2_optimality}, and (ii) in \ref{theorem:anm_v2_optimality} prove that $\m{x} = \sum_{i = 1}^L c_i\m{a}_i$ satisfies the two conditions in 
    Lemma \ref{lemma: optimality} and therefore solve 
    the original \ac{AST} problem (\ref{eq:anm}). Moreover, because of (\ref{eq:dual_support}),
    the objective value in (\ref{eq:anm_v2}) produced by $\mathcal{S}$ is the same as 
    the objective value produced by $\m{x} = \sum_{i = 1}^L c_i\m{a}_i$ in (\ref{eq:anm}). According 
    to Theorem \ref{theorem:anm_equivalence}, the value is the minimal objective 
    of (\ref{eq:anm_v2}). This concludes the proof for the backward statement.

\end{IEEEproof}
}

{\color{black}Theorem \ref{theorem:anm_v2_optimality} is important as it states the sufficient condition 
for finding solutions of (\ref{eq:anm_v2}). The condition plays an important role in the algorithmic 
design. An underlying fact behind Theorem \ref{theorem:anm_v2_optimality} 
is that (\ref{eq:anm_v2}) has infinitely many solutions, while they all produce the same 
objective value as well as the same dual certificate of support $\m{y}_r$. 
The next section introduces an iterative algorithm to solve (\ref{eq:anm_v2}) based on 
the condition}. Effectively, the algorithm simultaneously solves the original {\color{black} \ac{AST}}
problem (\ref{eq:anm}).

\section{A Coordinate Descent Method for \ac{AST}}\label{sec:algorithm}

{\color{black} This section discusses the design of the proposed coordinate descent solver for solving (\ref{eq:anm_v2}). 
The solver is designed to iteratively select a set of tuples $\mathcal{S}$ towards satisfying the 
conditions in Theorem \ref{theorem:anm_v2_optimality}. To ensure convergence, in each iteration the tuples 
in the set $\mathcal{S}$ are modified such that the objective value in the current iteration 
is smaller than in the previous one. In the following, we first discuss the specific 
steps of the solver, then explain details of its implementation, and lastly introduce a generic acceleration technique.} 

{\color{black}
\subsection{Descent Steps and Algorithmic Design}
}

Theorem \ref{theorem:anm_v2_optimality} reveals a simple but critical 
insight for solving \ac{AST}: the global optimal solution is found once a 
set of tuples $\mathcal{S}$ is properly chosen such that the two conditions of being a dual 
certificate are satisfied by its residual. The remaining question is then how to
 choose such a set of tuples $(c, \m{a})$. Inspired by previous work on applying 
 the coordinate descent to (\ref{eq:lasso}) \cite{friedman2010regularization}, 
 \cite{tibshirani2011solution}, a similar iterative approach is developed. {\color{black} The 
 key operation is to sequentially optimize for every tuple in $\mathcal{S}$ while keeping other tuples fixed. }

Suppose there are $L$ tuples in the set $\mathcal{S}$. The solver essentially repeats two kinds of operations:  
{\color{black}
\begin{itemize}
    \item{\textit{Refining}: The solver chooses a tuple $(c_i, \m{a}_i)$ 
    from the 
    current set $\mathcal{S}$. Let $\m{y}_r^{i} = \m{y}_r + c_i\m{a}_i$. 
    The tuple is refined with the following minimization: 
    \begin{align}\label{eq:tuple_coor_descent}
            (c_i, \m{a}_i) \leftarrow \underset{c \geq 0, \m{a}\in\mathcal{A}}{\mathrm{argmin}}\quad  
        \frac{\zeta}{2}\left\|\m{y}_r^i - c\m{a}\right\|_2^2 + c
    \end{align} 
    The residual $\m{y}_r$ is updated accordingly, 
    \begin{equation}\label{eq:residual_update1}
        \m{y}_r \leftarrow \m{y}_r^i - c_i\m{a}_i
    \end{equation}
    If in the result $c_i = 0$, the tuple is removed from the current set.
    }
    \item{\textit{Expanding}: The solver attempts to add a new tuple $(c_{L+1}, \m{a}_{L+1})$ 
    to $\mathcal{S}$. The tuple is 
    obtained with the following minimization: 
    \begin{align}\label{eq:new_tuple_descent}
            (c_{L+1}, \m{a}_{L+1}) 
            \leftarrow \underset{c \geq 0, \m{a}\in\mathcal{A}}{\mathrm{argmin}}\quad  
        \frac{\zeta}{2}\left\|\m{y}_r - c\m{a}\right\|_2^2 + c
    \end{align} 
    The residual $\m{y}_r$ is updated accordingly if $c_{L+1} > 0$, 
    \begin{equation}\label{eq:residual_update2}
        \m{y}_r \leftarrow \m{y}_r - c_{L+1}\m{a}_{L+1}
    \end{equation}
    }
\end{itemize}
}

\bluetext{In general, both (\ref{eq:tuple_coor_descent}) and (\ref{eq:new_tuple_descent}) are conic projections with shrinkage and thresholding. Their solutions are explained in the next subsection.}
The solver itself is essentially a finite state machine. It functions according to the current state of $\mathcal{S}$ 
as stated below: 
\begin{enumerate}
    \item Check whether $\mathcal{S}$ satisfies (ii) in Theorem \ref{theorem:anm_v2_optimality}. If not, perform 
    the refining operation and stay in 1). Otherwise, go to 2).
    \item Check whether $\mathcal{S}$ satisfies (i) in Theorem \ref{theorem:anm_v2_optimality}. If not, perform 
    the expanding operation and go back to 1). Otherwise, go to 3). 
    \item Return $\mathcal{S}$ as the solution to (\ref{eq:anm_v2}) as it reaches the optimal objective value. Exit.     
\end{enumerate}

The procedure above does not {\color{black} specify how to choose one tuple from the set when refining, which can be customized to be 
greedy, cyclic, or random, etc.} 
As an example, a pseudocode for using a cyclic sampling strategy with 
an initially empty set is given in \textbf{Algorithm} \ref{alg:cyclic}. {\color{black} Since (ii) in Theorem \ref{theorem:anm_v2_optimality}
is a strict equality, using it as the exit condition only yields solutions with tolerance $\varepsilon$ smaller
than machine precision. This is not always necessary. Therefore, in algorithm \ref{alg:cyclic}, a tolerance $\varepsilon$ 
is added, and
the condition is changed to be the absolute error between both sides of (ii) (in line 13). This condition  
also characterizes the convergence of algorithm \ref{alg:cyclic} as proved in the following theorem. 
\begin{theorem}\label{theorem:cyclic_convergence}
    Suppose $\mathcal{A}$ is symmetric with respect to the origin. For a 
    given $\m{y}$ and $\varepsilon >  0$, there exists $K < \infty$ such that within $K$ iterations,  
    algorithm \ref{alg:cyclic} 
    returns a set $\mathcal{S}$ whose objective value:  
    \begin{equation}\label{eq:anm_v2_obj_v2}
        h(\m{y}, \zeta, \m{\mathcal{S}}) = \frac{\zeta}{2}\left\|\m{y} - \sum\limits_{(c, \m{a})\in\mathcal{S}}
        c\m{a}\right\|_2^2 + \sum\limits_{(c, \m{a}) \in \mathcal{S}} c
    \end{equation}
    is at most 
    $\varepsilon$ larger than the global minimal objective of (\ref{eq:anm_v2}).      
\end{theorem}
\begin{IEEEproof}
    The proof first shows that algorithm \ref{alg:cyclic} exits with finitely many iterations. 
    Let $\mathcal{S}_k$ be the set of tuples in the $k$-th iteration 
    of algorithm \ref{alg:cyclic}.
    The proof considers three sequences with respect to $k = 0, 1,2, \ldots...$, 
    \begin{itemize}
        \item Sequence of objective: $h(\m{y}, \zeta', \mathcal{S}_k)$.
        \item Sequence of $h'(\m{y}, \zeta, \mathcal{S}_k)$: 
        \begin{equation}\label{eq:duality_gap}
            h'(\m{y}, \zeta, \mathcal{S}_k) = 
            \left|\zeta\left<\m{y}_r, \m{y} - \m{y}_r\right> - \sum\limits_{(c, \m{a}) \in \mathcal{S}_k} c\right|
        \end{equation}
        in which $\m{y}_r = \m{y} - \sum_{(c, \m{a})\in\mathcal{S}_k}
        c\m{a}$ and $\zeta' = \zeta/(1 - \delta)$ 
        is defined as in the initialization of algorithm \ref{alg:cyclic}, as well as $\delta = \varepsilon/(\zeta\left\|\m{y}\right\|_2^2 + \varepsilon)$.
        \item Sequence of $h''(\m{y}, \mathcal{S}_k)$:
        \begin{equation}\label{eq:dual_atomic_norm}
            h''(\m{y}, \mathcal{S}_k) = \mathrm{sup}_{a\in\mathcal{A}} \left<\m{y}_r, \m{a}\right>
        \end{equation}
        in which $\m{y}_r = \m{y} - \sum_{(c, \m{a})\in\mathcal{S}_k}
        c\m{a}$.
    \end{itemize}
    
    Algorithm \ref{alg:cyclic} replaces $\zeta$ with $\zeta' = \zeta/(1 - \delta)$ in minimization problems.
    Based on the minimization (line 7 and line 15), 
    the sequence $h(\m{y}, \zeta', \mathcal{S}_0), h(\m{y}, \zeta', \mathcal{S}_1), ...$ is a 
    bounded and monotonically decreasing sequence. Therefore, the sequence converges as 
    $k\rightarrow \infty$.
    
    The convergence of $h(\m{y}, \zeta', \mathcal{S}_k)$ means that: 
    \begin{equation}\label{eq:convergence_dual_norm}
        \lim\limits_{k \rightarrow \infty} h''(\m{y}, \mathcal{S}_k) \leq 1/\zeta' < 1/\zeta
    \end{equation}
    Otherwise $\lim\limits_{k\rightarrow \infty } h(\m{y}, \zeta', \mathcal{S}_k)$ would be 
    unbounded because of the expanding operation. A direct consequence of 
    (\ref{eq:convergence_dual_norm}) is that there exists $K_1 < \infty$, such that 
    $\forall k \geq K_1, h''(\m{y}, \mathcal{S}_k) < 1/\zeta $. 

    The convergence of $h(\m{y}, \zeta', \mathcal{S}_k)$ also means that in the limit, 
    every tuple in 
    $\mathcal{S}_k$ is stationary. Reusing the argument in (\ref{eq:anm_v2_deri_c}) to 
    (\ref{eq:anm_condition_a}) 
    with $\zeta' = \zeta/(1 - \delta)$ yields:   
    \begin{align}\nonumber
        \lim\limits_{k\rightarrow \infty} 
        h'(\m{y},\zeta, \mathcal{S}_k) & = \lim\limits_{k\rightarrow \infty}
        \sum_{(c, \m{a}) \in \mathcal{S}_k} c\left(1 - \zeta\left<\m{y}_r, \m{a}\right>\right) \\ \nonumber
        & =  \lim\limits_{k\rightarrow \infty} \sum_{(c, \m{a}) \in \mathcal{S}_k} c\left(1 - \zeta/\zeta'\right)\\ \label{eq:limit_i}
        & =   \lim\limits_{k\rightarrow \infty} \delta \sum_{(c, \m{a}) \in \mathcal{S}_k} c
    \end{align}
    The right-hand-side of (\ref{eq:limit_i}) is 
    upper bounded by the initial objective $h(\m{y}, \zeta', \mathcal{S}_0) = \frac{\zeta'}{2}\left\|\m{y}\right\|_2^2$. 
    With the definition of $\delta = \varepsilon/(\zeta\left\|\m{y}\right\|_2^2 + \varepsilon)$, there is further: 
    \begin{align}\nonumber
        \lim\limits_{k\rightarrow \infty} 
        h'(\m{y},\zeta, \mathcal{S}_k) & =
        \delta \sum_{(c, \m{a}) \in \mathcal{S}} c \\ \nonumber
        & \leq \frac{\delta}{2(1 - \delta)}\zeta\left\|\m{y}\right\|_2^2\\ \label{eq:delta_threshold} 
        & = \frac{\varepsilon}{2} \\ \nonumber
        & < \varepsilon 
    \end{align}
    Similarly, (\ref{eq:delta_threshold}) ensures that there exists $K_2 < \infty$ such that: 
    $\forall k \geq K_2, h'(\m{y}, \zeta, \mathcal{S}_k) < \varepsilon $. 

    Finally, according to the conditions in line 13 and line 14 of algorithm \ref{alg:cyclic}, it  
    returns a set $\mathcal{S}_k$ and its residual $\m{y}_r$ within $K = \mathrm{max}(K_1, K_2)$ iterations.
    With both $K_1$ and $K_2$ being finite, $K < \infty$.  

    The second part of the proof discusses the difference between $h(\m{y}, \zeta, \mathcal{S}_k)$ and 
    the global minimal objective of (\ref{eq:anm_v2}) when $\mathcal{S}_k$ is the set returned. 
    According to the theory of convex optimization, the optimal objective of 
    (\ref{eq:anm}) is lower bounded by that of its dual maximization problem \cite{boyd2004convex}. Since 
    (\ref{eq:anm}) and (\ref{eq:anm_v2}) share the same global minimal objective, the lower bound 
    applies to (\ref{eq:anm_v2}) as well. 
    
    The dual problem of (\ref{eq:anm}) \cite{ast}, \cite{hansen2019fast} 
    is : 
    \begin{equation}\label{eq:anm_dual}
        \begin{split}
          \underset{\m{y}_r'}{\mathrm{Minimize}} & \quad 
          \zeta\left<\m{y}_r', \m{y}\right> - \frac{\zeta}{2}\left\|\m{y}_r'\right\|_2^2\\
          \mathrm{subject\ to} & \quad \mathrm{sup}_{a\in\mathcal{A}} \left<\m{y}_r', \m{a}\right> \leq 1/\zeta
        \end{split}
    \end{equation}
    Since the residual $\m{y}_r$ of returned set is a feasible for (\ref{eq:anm_dual}), its 
    objective value can also be used to bound the difference.
    Therefore, the difference between $h(\m{y},\zeta,\mathcal{S}_k)$ and the global minimal objective 
    of (\ref{eq:anm_v2}) is upper bounded by:
    \begin{equation}
        h(\m{y},\zeta,\mathcal{S}_k)
     - \zeta\left<\m{y}_r, \m{y}\right> + \frac{\zeta}{2}\left\|\m{y}_r\right\|_2^2
     \leq h'(\m{y}, \zeta, \mathcal{S}_k)
    \end{equation}
    which is further upper bounded by $\varepsilon$. This concludes the proof.

\end{IEEEproof}

}

\begin{algorithm}[htbp!]
\caption{Cyclic  Coordinate Descent for \ac{AST}}\label{alg:cyclic}
\begin{algorithmic}[1]
\STATE {\textbf{Input}:} \\
    \hspace{.5em} Observation vector $\m{y}$;\ Atomic set $\mathcal{A}$;\  Threshold $\zeta$;\\
    \hspace{.5em} Tolerance $\mathrm{\varepsilon}$; Maximum Iteration $K$; 
\STATE {\textbf{Initialize}:} \\
    \hspace{.5em} Empty set of tuples $\mathcal{S}$;\ Residual vector $\m{y}_r \leftarrow \m{y}$;\\
    \hspace{.5em} $L \leftarrow 0, i \leftarrow 1$; {\color{black}$\delta \leftarrow \varepsilon/(\zeta\left\|\m{y}\right\|_2^2 + \varepsilon)$
    ; $\zeta' \leftarrow \zeta/(1 - \delta)$;} 
\FOR{$k = 1,2,.., K$}
    \IF{$i \leq L$}   
        \STATE{$(c_i, \m{a}_i) \leftarrow \left[\mathcal{S}\right]_i$} 
        \STATE{$\m{y}_r^i \leftarrow \m{y}_r + c_i\m{a}_i$}
        \STATE{ $(c_i, \m{a}_i) \leftarrow \underset{c \geq 0, \m{a}\in\mathcal{A}}{\mathrm{argmin}}\  
        {\color{black}\frac{\zeta'}{2}}\left\|\m{y}_r^i - c\m{a}\right\|_2^2 + c$}
         \STATE{$\m{y}_r \leftarrow \m{y}_r^i - c_i\m{a}_i$, $\left[\mathcal{S}\right]_i\leftarrow (c_i, \m{a}_i)$}
         \IF{$c_i == 0$} \STATE{Remove $(c_i, \m{a}_i)$ from $\mathcal{S}$, $L \leftarrow L - 1$, $i \leftarrow i - 1$}
         \ENDIF 
         \STATE{$i \leftarrow i + 1$} \vspace{0.05em}
    \ELSIF{$\left|\sum_{j}^{L} c_j - \zeta\left<\m{y}_r, \m{y} - \m{y}_r\right>\right| \leq \mathrm{\varepsilon}$}\vspace{0.05em}
        \IF{$\sup_{a\in\mathcal{A}}\left<\m{y}_r, \m{a}\right> > {\zeta}^{-1}$}\vspace{0.05em}
            \STATE{$ (c_{L+1}, \m{a}_{L+1}) 
            \leftarrow \underset{c \geq 0, \m{a}\in\mathcal{A}}{\mathrm{argmin}}\  
            {\color{black}\frac{\zeta'}{2}}\left\|\m{y}_r - c\m{a}\right\|_2^2 + c$}
            \STATE{$\m{y}_r \leftarrow \m{y}_r - c_{L+1}\m{a}_{L+1}$}
            \STATE{Add $(c_{L+1}, \m{a}_{L+1})$ to $\mathcal{S}$, $L \leftarrow \left|\mathcal{S}\right|, i \leftarrow 1$}
        \ELSE
            \STATE{\textbf{Break}}
        \ENDIF
    \ELSE
        \STATE{$i \leftarrow 1$}
    \ENDIF
\ENDFOR
\STATE \textbf{return} The set of tuples $\mathcal{S}$, Residual $\m{y}_r$
\end{algorithmic}
\end{algorithm}

{\color{black}
Theorem \ref{theorem:cyclic_convergence} provides theoretical support for the convergence 
of algorithm \ref{alg:cyclic}. An interesting observation is that it doesn't account for the 
case $\varepsilon = 0$. This is basically because for $\varepsilon = 0$, $\zeta' = \zeta$. And 
it's possible to have a sequence 
$h''(\m{y}, \mathcal{S}_k)$ converge to $1/\zeta$ in the limit but never satisfies 
$h''(\m{y}, \mathcal{S}_k) \leq 1/\zeta$ for any finite $k$, which means algorithm \ref{alg:cyclic} 
would never terminate unless $k$ reaches the maximum number of iterations. Though it might take an infinite number of iterations, with $\varepsilon = 0$ the two conditions in Theorem \ref{theorem:anm_v2_optimality} will be satisfied by $\mathcal{S}_k$ in the limit.
}

\begin{remark}\label{remark:ambiguity}
    So far in algorithm \ref{alg:cyclic}, it is not required that 
    different tuples from $\mathcal{S}$ must have different elements 
    from $\mathcal{A}$. Although the solution $\m{x}$ to (\ref{eq:anm}) is unique, 
    the optimal set that solves (\ref{eq:anm_v2}) is not unique unless it is restricted 
    that different tuples must have different elements from $\mathcal{A}$, 
    i.e., $\m{a}_i \neq \m{a}_j$ if $i \neq j$. Therefore, upon {\color{black}termination} 
    of algorithm \ref{alg:cyclic}, multiple tuples in $\mathcal{S}_k$ 
    might have the same element from $\mathcal{A}$. Such ambiguities do not prevent the 
    algorithm from {\color{black}termination as suggested in Theorem \ref{theorem:cyclic_convergence}}.
\end{remark}
\vspace{-.5em}
\bluetext{
\subsection{Conic Projection and Implementation}
}

\bluetext{
This subsection addresses the implementation of the proposed coordinate descent solver, as well as the 
solution of conic projection with shrinkage and thresholding. 
}

Besides its cyclic sampling strategy, algorithm \ref{alg:cyclic} exemplifies 
several key features of the proposed coordinate descent framework for {\color{black} \ac{AST}}. 
Throughout the iterations, only the set of tuples $\mathcal{S}$, the residual vector $\m{y}_r$, 
and the original vector $\m{y}$ are being stored. 
In each iteration the relatively expensive steps are {\color{black} refining (line 7),  expanding (line 15), 
or checking the condition (i) (line 14)}. All other steps have only $\mathcal{O}(N)$ 
computational complexity. For the classical atomic set (\ref{eq:atomic_set_dft}),
these steps have only $\mathcal{O}(N\mathrm{log} N)$ operations. 

{\color{black} The implementation of algorithm \ref{alg:cyclic} requires a reliable way 
of solving (\ref{eq:tuple_coor_descent}) and (\ref{eq:new_tuple_descent}), which correspond 
to refining and expanding, respectively}. \bluetext{Both problems can be treated as conic projections. Since the set $\mathcal{A}$ is not necessarily convex, projecting onto $\mathrm{cone}(\mathcal{A})$ is not a convex problem. Nonetheless, the projection still has a separable structure. For instance, consider the following derivation based on the objective in (\ref{eq:new_tuple_descent}): 
\begin{equation}\label{eq:conic_proj}
\begin{split}
    \frac{\zeta}{2} & \left\|\m{y}_r - c\m{a}\right\|_2^2 + c = \frac{\zeta}{2}\left[\left\|\m{y}_r\right\|_2^2 + \right. \\ 
    & \left. \left(c\left\|\m{a}\right\|_2 + \frac{1/\zeta - \left<\m{y}_r, \m{a}\right>}{\left\|\m{a}\right\|_2} \right)^2 - 
    \frac{(1/\zeta - \left<\m{y}_r, \m{a}\right>)^2}{ \left\|\m{a}\right\|_2^2}\right]
\end{split}
\end{equation}
Based on (\ref{eq:conic_proj}), the solution to (\ref{eq:new_tuple_descent}) is readily calculated as: 
\begin{align}\label{eq:conic_proj_sol_a}
    \m{a}^* & = \mathrm{argmax}_{\m{a}\in\mathcal{A}} \quad  \frac{1}{\left\|\m{a}\right\|_2}\left(
    \left<\m{y}_r, \m{a}\right> - 1/\zeta\right)\\ \label{eq:conic_proj_sol_c}
    c^* & = \left\{\begin{matrix}
        0, & \ \left<\m{y}_r, \m{a}^*\right> \leq \frac{1}{\zeta} \\[.2em]
      \frac{1}{\left\|\m{a}^*\right\|_2^2}\left(
    \left<\m{y}_r, \m{a}^*\right> - 1/\zeta\right), & \ \left<\m{y}_r, \m{a}^*\right> > \frac{1}{\zeta} \\
    \end{matrix}\right.
\end{align}
It's clear that as long as $\m{a}^*$ can be calculated, (\ref{eq:new_tuple_descent}) and (\ref{eq:tuple_coor_descent}) can be solved. 
The shrinkage and thresholding is related to the threshold $1/\zeta$ as reflected in (\ref{eq:conic_proj_sol_c}).
}

In general, 
the complexity of solving (\ref{eq:tuple_coor_descent}) or 
(\ref{eq:new_tuple_descent}) as well as calculating 
$\sup_{a\in\mathcal{A}}\left<\m{y}_r, \m{a}\right>$ depends on the structure 
of $\mathcal{A}$. The rest of this section addresses these operations for 
$\mathcal{A}$ defined as in (\ref{eq:atomic_set_dft}). 
The calculation can be generalized to atomic sets in (\ref{eq:atomic_set_dft_2d}), 
(\ref{eq:atomic_set_dft_mmv}). With $\mathcal{A}$ defined in (\ref{eq:atomic_set_dft}), 
the vector space of interests is $\mathbb{C}^N$, 
and the inner-product is defined as $\left<\m{x},\m{y}\right> = \mathrm{Re}\left\{\m{x}^\hermitian \m{y}\right\}$.    
With the derivation in \bluetext{(\ref{eq:conic_proj})}, (\ref{eq:new_tuple_descent}) under (\ref{eq:atomic_set_dft}) is equivalent to the 
 following optimization problem: 
\begin{align}\label{eq:obj_single_atom}
          \underset{\phi, f, c}{\mathrm{Minimize}} & \quad 
          c\left(1 - \zeta\left<\m{y}_r, e^{\mathrm{1j}
          \phi}\m{a}(f)\right>\right) + \frac{Nc^2\zeta}{2} \\ \nonumber
          \mathrm{subject\ to} & \quad 
          c \geq 0; \ \phi, f\in \left[0, 2\pi\right);
\end{align}
Let $c^*, f^*, \phi^*$ be the solution. The problem has a separable structure: 
\begin{align}\label{eq:opt_f}
    f^* & = \mathrm{argmax}_f \left|\m{y}_r^\hermitian\m{a}(f)\right| \\ \label{eq:opt_c}
    c^* & = \left\{\begin{matrix}
        0, & \ \left|\m{y}_r^\hermitian\m{a}(f^*)\right| \leq \frac{1}{\zeta} \\
      \frac{1}{N}\left(\left|\m{y}_r^\hermitian\m{a}(f^*)\right| - \frac{1}{\zeta}\right), & \ \left|\m{y}_r^\hermitian\m{a}(f^*)\right| > \frac{1}{\zeta} \\
    \end{matrix}\right.\\  \label{eq:opt_phi}
    \phi^* & = -\angle\left(\m{y}_r^\hermitian\m{a}(f^*)\right)  
\end{align}

Among the three, the key step is (\ref{eq:opt_f}) \bluetext{ which corresponds to (\ref{eq:conic_proj_sol_a})}. Both $c^*$ and $\phi^*$ depend on $f^*$. 
The problem is essentially locating the maximum of a 
polynomial on the unit circle. For this purpose, 
the low-complexity approach in \ac{NOMP} \cite{newtonized} can be employed as specified 
in the following algorithm \ref{alg:newtonized}:

\begin{algorithm}[htbp!]
\caption{Calculating the Maximum on the Unit Circle}\label{alg:newtonized}
\begin{algorithmic}[1]
\STATE {\textbf{Input}:} \\
    \hspace{.5em} Complex vector $\m{y} \in \mathbb{C}^N$;\ Tolerance $\mathrm{tol} = 10^{-12}$; \\ \hspace{.5em} Oversampling Ratio $r = 16$\\  
\STATE {\textbf{Initialize}:} \\
    \hspace{.5em} Construct $\hat{\m{y}} \in \mathbb{C}^{2N}$ such that:\\
    \hspace{.5em} $\left[\hat{\m{y}}\right]_{1:N} = \m{y}, \left[\hat{\m{y}}\right]_{N+1:2N} = \m{0}$ \vspace{0.2em}
\STATE \bluetext{ $\tilde{\m{y}} \leftarrow \left[\mathcal{F}^{-1}\left\{\left|\mathcal{F}\left\{\hat{\m{y}}\right\}\right|^2\right\}\right]_{1:N}$ } \vspace{0.2em}
\STATE {Evaluate $\mathrm{Re}\left\{\tilde{\m{y}}^\hermitian\m{a}(f)\right\}$ on a uniform grid of $rN$ points $f = 0, \frac{2\pi}{rN}, ..., \frac{2\pi(rN - 1)}{rN}$ using \ac{FFT}}
\STATE {$f^* \leftarrow $ the on-grid $f$ with the largest 
$\mathrm{Re}\left\{\tilde{\m{y}}^\hermitian\m{a}(f)\right\}$ among these $rN$ points}
\WHILE{ \textbf{True}}
    \STATE{$\Delta f \leftarrow \mathrm{Re}\left\{\tilde{\m{y}}^\hermitian \nabla_f \m{a}(f^*)\right\}/ \mathrm{Re}\left\{\tilde{\m{y}}^\hermitian \nabla_f^2  \m{a}(f^*)\right\}$}
    \IF{$\Delta f \leq \mathrm{tol}$}
        \STATE{\textbf{Break}}
    \ENDIF
    \STATE{$ f^* \leftarrow f^* - \Delta f$}
\ENDWHILE
\STATE \textbf{return} $f^*$ such that $\left|\m{y}^\hermitian\m{a}(f^*)\right|^2 = \sup_f \left|\m{y}^\hermitian\m{a}(f)\right|^2$
\end{algorithmic}
\end{algorithm}

With \ac{FFT}, algorithm \ref{alg:newtonized} has 
$\mathcal{O}\left(N\mathrm{log} N\right)$ operations. Specifically, the initialization 
steps line 2 and line 3 calculate $\tilde{\m{y}}$ such that 
$\mathrm{Re}\left\{\tilde{\m{y}}^\hermitian\m{a}(f)\right\} = \left|\m{y}^\hermitian\m{a}(f)\right|^2$. 
Line 4 and line 5 then perform initial over sampling on a uniform grid. Line 6 to line 12 
have only $\mathcal{O}(N)$ operations. These steps use Newton's method to calculate the 
off-grid maximum of the function $\mathrm{Re}\left\{\tilde{\m{y}}^\hermitian\m{a}(f)\right\}$. 
It has been shown previously that $r > 1$ is necessary to ensure the convergence to the true 
maximum as the function $\mathrm{Re}\left\{\tilde{\m{y}}^\hermitian\m{a}(f)\right\}$ on 
the interval $[0, 2\pi)$ is non-convex \cite{newtonized}. The default value $r = 16$ is 
established empirically
\footnote{A rigorous discussion on how large $r$ should be is related to the spacing 
between roots of a polynomial on the unit circle, which is beyond the scope.} 
and is found to be sufficient in this work. Algorithm \ref{alg:newtonized} provides 
a low-complexity method to evaluate $h''(\m{y}, \mathcal{S})$. Together, with (\ref{eq:opt_f}) - (\ref{eq:opt_phi}), 
it provides a solution to (\ref{eq:tuple_coor_descent}) or (\ref{eq:new_tuple_descent}). 
With the missing pieces provided by algorithm \ref{alg:newtonized}, algorithm 
\ref{alg:cyclic} is readily applicable to {\color{black} \ac{AST}} problems with the atomic set (\ref{eq:atomic_set_dft}). 
{\color{black} The major hyperparameter in algorithm \ref{alg:cyclic} is the tolerance $\varepsilon$. A heuristic way of setting $\varepsilon$ is to set 
$\varepsilon = 10^{-2}\zeta\left\|\m{y}\right\|_2^2/(1 - 10^{-2})$ such that \bluetext{ $1/\zeta' = (1 - 10^{-2})/\zeta$ }, as setting small $\varepsilon$ with large $\zeta\left\|\m{y}\right\|_2^2$ often leads to slow convergence. $\epsilon$ can also be set to a specific value (e.g., $\varepsilon = 10^{-12}$) to bound the distance to then minimal objective when necessary.}

\begin{figure*}[htbp!]
\centering
\subfloat[$0$-th Iteration]{\includegraphics[width=.3\linewidth]{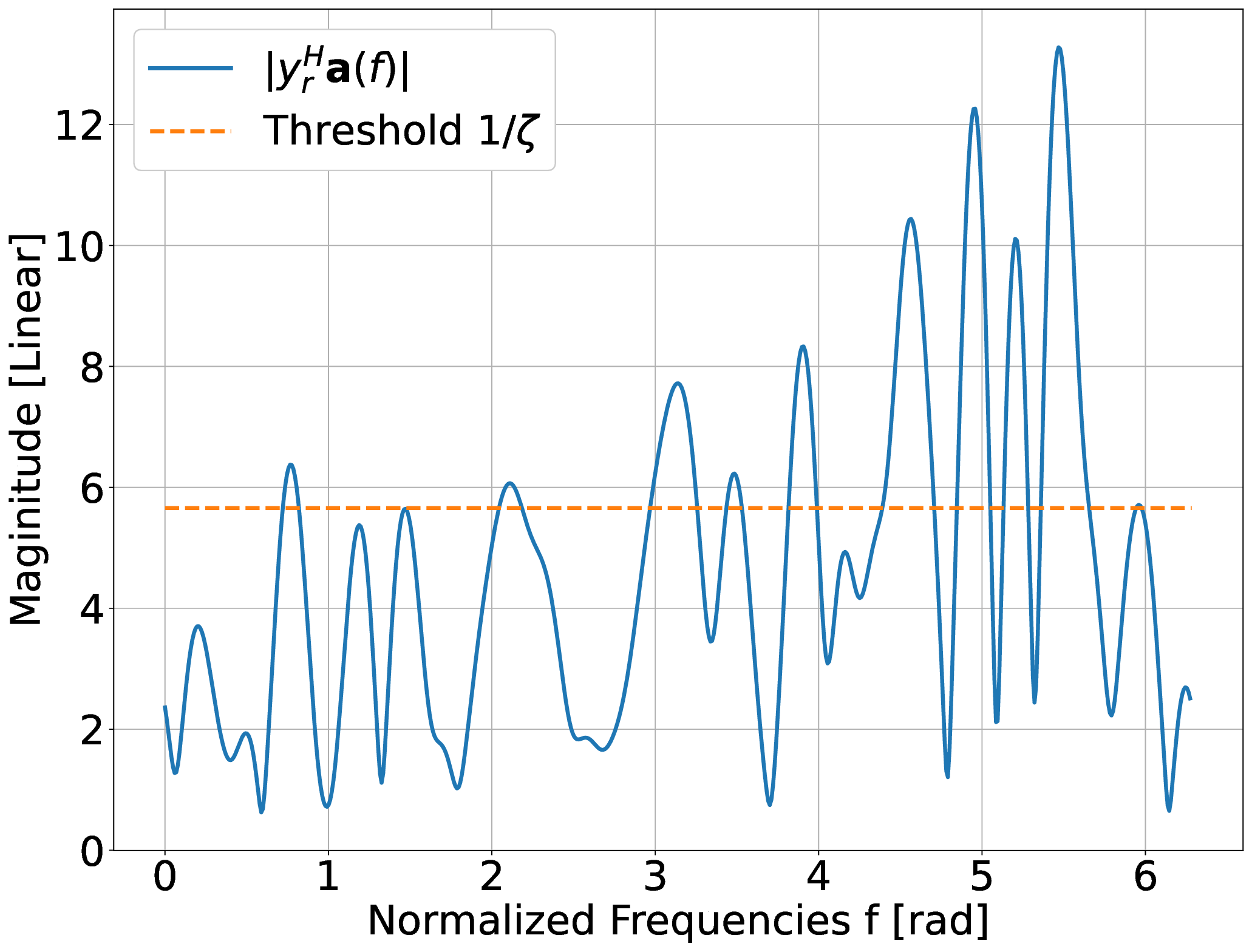}%
\label{fig:visual_randomized_0_iter}}
\hfil
\subfloat[$10$-th Iteration]{\includegraphics[width=.3\linewidth]{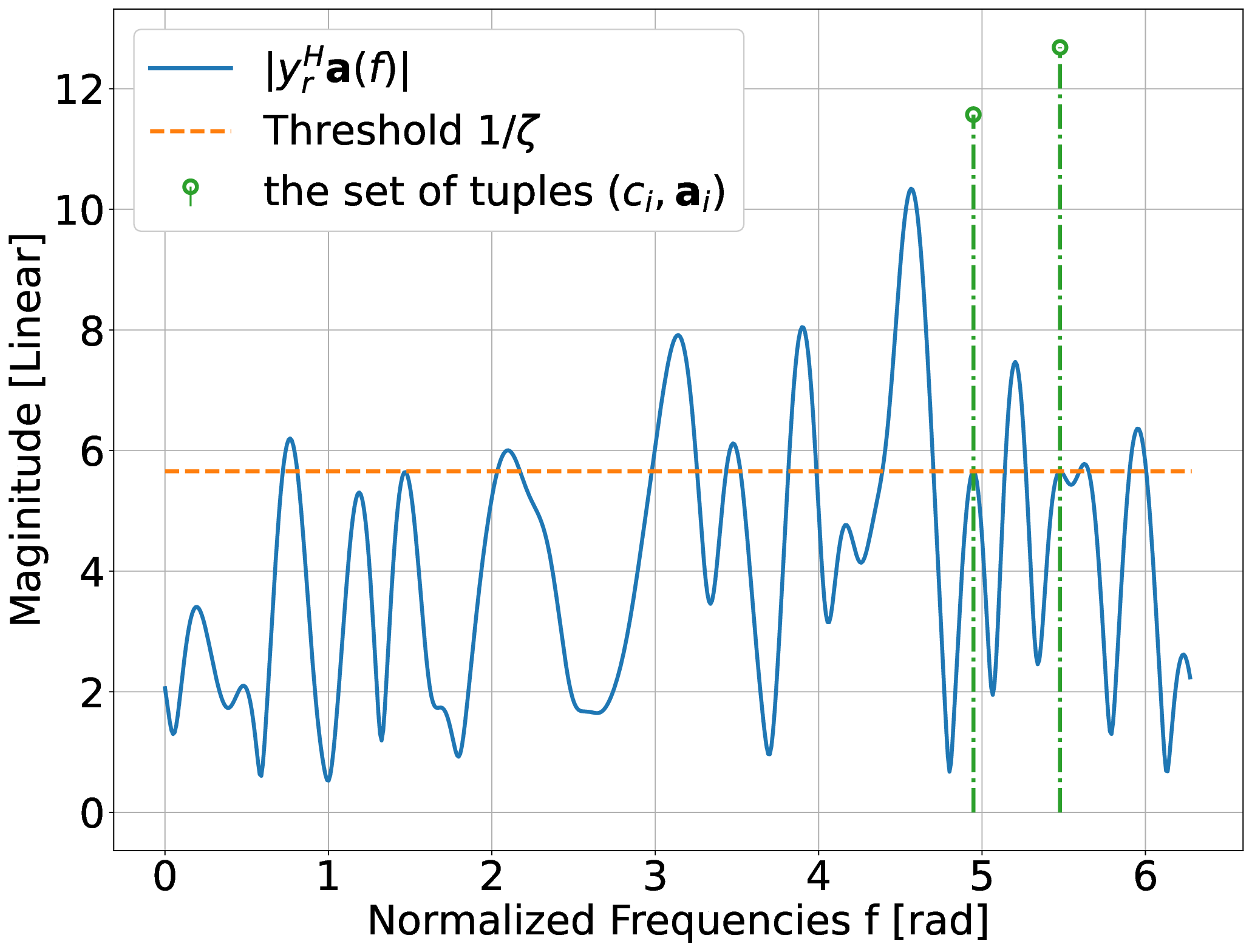}%
\label{fig:visual_randomized_10_iter}}
\hfil 
\subfloat[{\color{black} Termination}]{\includegraphics[width=.3\linewidth]{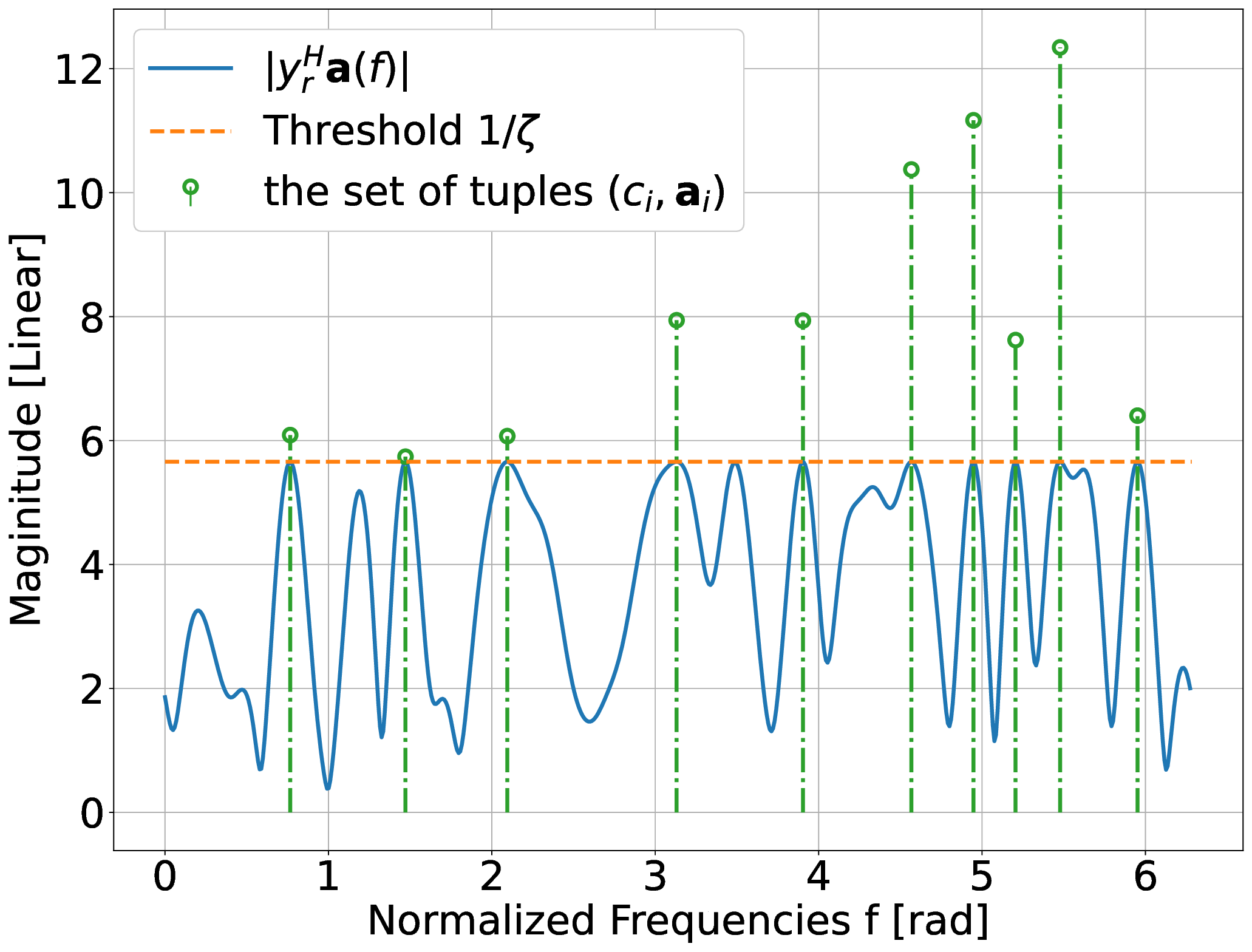}%
\label{fig:visual_randomized_converged}}
\caption{Visualization of convergence in cyclic coordinate descent. Through iterations, tuples $(c_i, \m{a}_i)$ are added to the set $\mathcal{S}$. Each {\color{black} green} stem in the plot represents an element $\m{a}_i$ with corresponding scaling $c_i + 1/\zeta$ in the tuple. With 10 iterations, only 2 tuples are added to $\mathcal{S}$. At {\color{black}termination, the solver returned 10 tuples in $\mathcal{S}$}.}
\label{fig:visual_randomized}
\end{figure*}

Figure \ref{fig:visual_randomized} provides visualizations for the {\color{black} expanding and refining steps} of algorithm \ref{alg:cyclic} on an exemplary problem in which $N = 32, \m{y} \in \mathbb{C}^{32}$. 
Each entry $\left[\m{y}\right]_i$ is independently sampled from $\mathcal{CN}\left(0, 1\right)$. 
The parameter $\zeta$ is set to $\zeta = 1/\sqrt{N} = 1/\sqrt{32}$.  
The algorithm {\color{black} terminates within 400 iterations with $\varepsilon = 10^{-12}$}.

\subsection{A Generic Acceleration Technique} \label{sec:solution_path}

Before evaluating  the performance of the proposed algorithm, this section
introduces a generic technique that can speed up the convergence of the
{\color{black}solver}. The major weakness of algorithm \ref{alg:cyclic} is 
the rate of convergence. In previous works on block coordinate descent 
algorithms applied to \ac{LASSO} problems,
a linear convergence rate has been established \cite{li2017faster}, 
as the problem (\ref{eq:lasso}) is convex. 
The linear rate of convergence for \ac{LASSO} is observed numerically even 
when matrix $\m{A}$ is badly conditioned. Otherwise, for general non-smooth {\color{black} but convex}
problems, the worst-case rate of convergence of coordinate descent is sublinear 
\cite{Beck2013OnTC}.  

From a theoretical perspective, it is hard to establish similar results for the mixed integer problem (\ref{eq:anm_v2}), as {\color{black} (\ref{eq:anm_v2}) is non-convex and} the number of coordinates $L$ is changing throughout iterations. From a practical perspective, a linear convergence rate is only observed when the 
set {\color{black}$\mathcal{S}_k$ is nearly optimal and no more expanding operation is needed}. At this stage, (\ref{eq:anm_v2}) becomes very similar to the \ac{LASSO} problem (\ref{eq:lasso}) as in each tuple $(c_i, \m{a}_i)$ the element $\m{a}_i$ is almost 
stationary. Therefore, the technique in this section is designed generically to quickly get to the stage that has a linear rate of convergence. In algorithm \ref{alg:cyclic}, a new tuple would only be added if 
{\color{black} the expanding operation (line 15) is performed}. {\color{black} The purpose of this design is to prevent the set $\mathcal{S}$ from growing too quickly. However, it also slows down convergence as repetitive refining steps yield only diminishing  benefits}. 

\begin{figure}[htbp!]
\centering
\includegraphics[width=.95\linewidth]{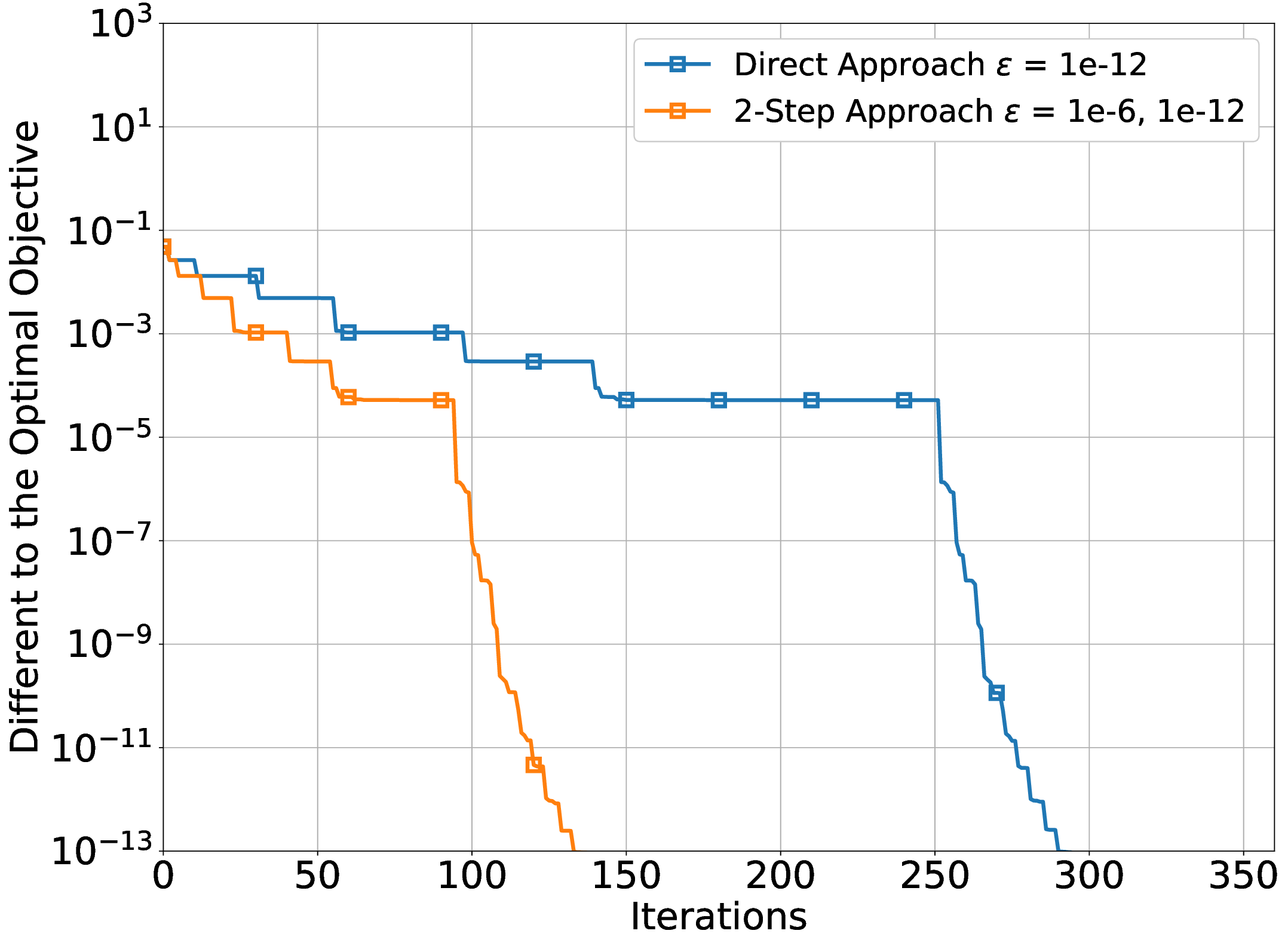}
\caption{Visualization of convergence in cyclic coordinate descent. The $y$-axis characterizes the difference to the optimal objective {\color{black} obtained by \ac{ADMM} }. The two-step strategy saves approximately half of the iterations.}
\label{fig:comparison_tolerance}
\end{figure}

The unnecessary iterations can be reduced by calling algorithm \ref{alg:cyclic}
twice in which the result from the first call is used to initialize the second call as a warm start. The first call uses a larger tolerance (e.g., {\color{black}$\varepsilon = 10^{-6}$}) and the second call uses the desired precision (e.g., {\color{black}$\varepsilon = 10^{-12}$}). The initialization steps (line 2) in algorithm \ref{alg:cyclic} can be altered to accommodate non-empty set $\mathcal{S}$. A comparison of the rates between the two-step approach and the direct approach is provided in figure 
\ref{fig:comparison_tolerance}. The problem being solved has the same $\zeta = 1/\sqrt{N} = 1/\sqrt{32}$ as that of figure \ref{fig:visual_randomized} but a different sampled vector $\m{y}$. As predicted, solving the problem with a relaxed tolerance first allows tuples to be added quickly to $\mathcal{S}$, after which a linear convergence rate is {\color{black} observed}. 

This concludes the discussion on algorithmic development of the proposed approach. 
In fact, the method can be {\color{black} further customized with the strategy of a solution path \cite{tibshirani2011solution}, or with a better heuristic that balances the number of the refining and expanding steps to terminate quickly, etc. Those explorations and theoretical analysis on the convergence of the approach are left as future work.} The next section provides numerical experiments that showcase the performances of the coordinate descent approach on various {\color{black}\ac{AST}}. 

\section{Numerical Experiments} 

\bluetext{ This section includes numerical examples on time trials for classic \ac{AST} \cite{ast}, two-dimensional \ac{MMV} \ac{AST} \cite{compressive2d}, separation-free weighted \ac{AST} \cite{yang2016enhancing}, and classic \ac{AST} in compressed scenarios}. \bluetext{ Due to the space limit, relevant derivations on  how the key step (\ref{eq:conic_proj_sol_a}) is evaluated for different atomic sets $\mathcal{A}$ is moved to appendix. In the subsections, the implementation for \ac{NOMP} used for comparisons was directly downloaded from authors' original repository \cite{nomp_software}.   }  


\subsection{Time Trials}\label{sec:tt}

{\color{black} Due to its low per-iteration complexity,  the proposed coordinate descent solver can solve \ac{AST} problems with large $N$ when \bluetext{ they have sparse structures}. For a specific \ac{AST} problem (\ref{eq:anm}) \bluetext{ in convex formulation}, its sparsity is qualitatively evaluated by $L_\mathrm{min}$, the minimum number of distinct elements needed to decompose its solution $\m{x}^*$ \footnote{{\color{black}$L_\mathrm{min}$ is also the minimum number of tuples in all solutions to (\ref{eq:anm_v2}). See also remark \ref{remark:ambiguity}.}}. An \ac{AST} problem is sparse if $L_\mathrm{min} \ll N$. The purpose of this subsection is to verify the efficiency of the solver on such occasions.}

{\color{black}To show that the solver is efficient, a sequence of sparse \ac{AST}} problems are constructed. {\color{black} The dimensions of the \ac{AST} problems include} $N = \{32, 64, 128, 256, 1024, 2048, 4096\}$ in which each entry of $\m{y} \in \mathbb{C}^N$ is sampled from circularly symmetric complex Gaussian distribution \bluetext{ $\mathcal{CN}(0 , 1)$}. The atomic set $\mathcal{A}$ is defined as in (\ref{eq:atomic_set_dft}). {\color{black} To control $L_\mathrm{min}$, the value $\zeta = \left(N\mathrm{log}{N/4}\right)^{-1/2}$ is used for $N$. }
Specifically in this experiment,  $\zeta = \left(N\mathrm{log}{N/4}\right)^{-1/2}$ yields $L_{\mathrm{min}} \leq 20$ for all $N$ with high probability.   

\begin{figure}[htbp!]
\centering  
\includegraphics[width=.95\linewidth]{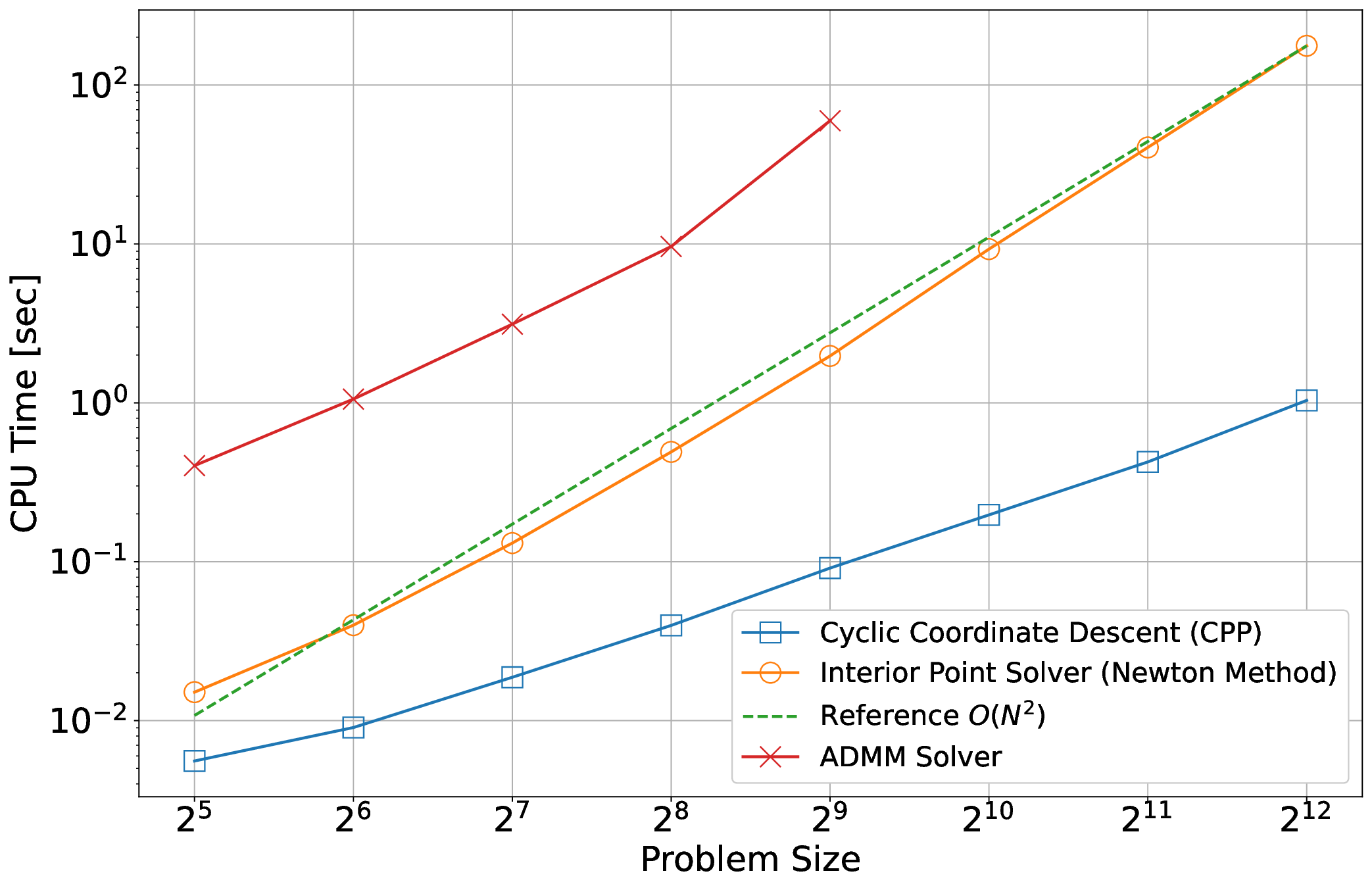}
\caption{Time Trials on sparse {\color{black} \ac{AST}} problems. }
\label{fig:comparison_timetrials}
\end{figure}

The \ac{ADMM} solver and the \ac{SOTA} interior-point solver from \cite{hansen2019fast} are employed for comparison. They're downloaded directly from the author's original repository on Github \cite{fast_ast_repo}. The mexfile in the implementation is then built and executed in MATLAB 2020b \cite{MATLAB}. On the other hand, the coordinate descent method is implemented in CPP with libraries FFTW v3.3.10 and Eigen v3.4.1 with a Python wrapper. The results of the time trials are provided in figure \ref{fig:comparison_timetrials}. For the interior point and the \ac{ADMM} solver, each data point is an average over 20 Monte Carlo trials. For the coordinate descent {\color{black}solver}, each point is an average over 200 trials.   

{\color{black} Empirically}, for sparse \ac{AST} problems, the interior-point solver with fast implementation for solving Toeplitz linear systems has $\mathcal{O}(N^2)$ complexity regardless of $L_{\mathrm{min}}$, whereas the complexity of the proposed method has \bluetext{$\mathcal{O}(L_{\mathrm{min}}^2 N\log(N))$} behavior. As indicated in figure \ref{fig:comparison_timetrials}, when $L_{\mathrm{min}} \ll N$, the proposed method has a clear advantage over the \ac{SOTA} interior-point method.

The fast computations on Toeplitz linear system which makes the \ac{SOTA} interior-point method behave like $\mathcal{O}(N^2)$ are only implemented for the simplest \ac{DFT} atomic set (\ref{eq:atomic_set_dft}). No other atomic set $\mathcal{A}$ was addressed in \cite{hansen2019fast}. In fact, using the \ac{SOTA} interior point method to solve more generalized \ac{AST} problems is complicated due to the evaluation of Hessian matrix. \bluetext{ On the other hand, the coordinate descent solver} can be generalized easily.

\subsection{Application to \ac{MMV} {\color{black} \ac{AST}} with 2D Frequencies}

It is well known that \ac{ANM} problems with \ac{MMV} (with atomic set (\ref{eq:atomic_set_dft_mmv})) or with two-dimensional frequencies (with atomic set (\ref{eq:atomic_set_dft_2d})) can be decoupled \cite{exactjoint}, \cite{decoupled17}. Decoupling is to formulate the \ac{SDP} constraint smartly such that the size of the problem is $N + M$ stead of $NM + 1$  in the case of (\ref{eq:atomic_set_dft_mmv}) with $M$ samples, or is $N + N$ instead of $N^2 + 1$ in the case of (\ref{eq:atomic_set_dft_2d}) with both \ac{DFT} vectors of length $N$. However, due to the limitation of the bounded real lemma, decoupling \ac{ANM} problems with two-dimensional \ac{DFT} set and \ac{MMV} is not a trivial task. To the best of our knowledge, \cite{decoupled2021} is the \ac{SOTA} along this direction. However, in \cite{decoupled2021} the authors resorted to the construction of a covariance matrix with redundancy reduction. Although this technique reduces the complexity of the problem, it also loses the robustness with respect to the correlation among signal sources. And to ensure the quality of the covariance construction, a large number of samples (for instance, $M \geq 200$) are needed. 

Without decoupling, the complexity of solving a two-dimensional \ac{MMV} \ac{ANM} problem in the \ac{SDP} formulation is about $\mathcal{O}((N^2 + M)^3)$. Such high complexity can be reduced with the proposed {\color{black}solver}. Let $\m{Y} \in \mathbb{C}^{N \times N \times M}$ be the tensor of interest. The two-dimensional \ac{MMV} {\color{black} \ac{AST}} problem can be equivalently formulated as: 
\begin{equation}\label{eq:anm_2D_MMV_ANM}
\begin{split}
  \underset{\begin{smallmatrix}
      L, f_i^1, f_i^2,\\
      \m{c}_i \in \mathbb{C}^{M} \\
  \end{smallmatrix}}{\mathrm{Minimize}} & \quad \sum\limits_{i}^L \left\|\m{c}_i\right\|_2 + \frac{\zeta}{2}\left\|\m{Y} - \sum\limits_{i}^L \m{a}(f_i^1) \otimes \m{a}(f_i^2) \otimes \m{c}_i \right\|_2^2\\
\end{split}
\end{equation}

 \bluetext{ Problem (\ref{eq:anm_2D_MMV_ANM}) can be solved by \textbf{Algorithm}  \ref{alg:cyclic}. with the following two-dimensional \ac{MMV} atomic set $\mathcal{A}$ (\ref{eq:atomic_set_dft2d_mmv}). 
\begin{equation}
  \label{eq:atomic_set_dft2d_mmv}
  \left\{\mathbf{a}\left(f^1\right) \otimes \m{a}\left(f^2\right) \otimes \hat{\m{c}} \left| \left[\m{a}\left(f\right)\right]_i = e^{\mathrm{1j}(i - 1)f}, \left\|\hat{\m{c}}\right\|_2 = 1\right.\right\}
\end{equation}
Note that in this case, $\m{a}(f) \in \mathbb{C}^N$ are \ac{DFT} vectors,  $\hat{\m{c}} \in \mathbb{C}^M$ are complex vectors with unit norm, and $\m{a} \in \mathcal{A}$ are tensors from the set. It's obvious that all elements from $\m{a} \in \mathcal{A}$ have the same Frobenius norm, i.e., $\left\|\m{a}\right\|_2 = N$. Therefore, (\ref{eq:conic_proj_sol_a}) and (\ref{eq:conic_proj_sol_c}) on a residual tensor $\m{Y}_r$ are adapted accordingly: 
\begin{align}\label{eq:conic_proj_sol_a_2dmmv}
    \m{a}^* & = \mathrm{argmax}_{f^1, f^2, \hat{\m{c}}^* } \quad  \frac{1}{N}\left(
    \left<\m{Y}_r, \m{a}\right> - 1/\zeta\right)\\ \label{eq:conic_proj_sol_c_2dmmv}
    c^* & = \left\{\begin{matrix}
        0, & \ \left<\m{Y}_r, \m{a}^*\right> \leq \frac{1}{\zeta} \\[.2em]
      \frac{1}{N^2}\left(
    \left<\m{Y}_r, \m{a}^*\right> - 1/\zeta\right), & \ \left<\m{Y}_r, \m{a}^*\right> > \frac{1}{\zeta} \\
    \end{matrix}\right.
\end{align}
 in which the inner-product between two tensors are defined as in (\ref{eq:tensor_inner_prod}).}  
 With two-dimensional $\ac{FFT}$, the complexity of \bluetext{ evaluating (\ref{eq:conic_proj_sol_a_2dmmv}) (one time per iteration)} is only $\mathcal{O}\left(M(N\log N)^2\right)$. \bluetext{ Details on solving (\ref{eq:conic_proj_sol_a_2dmmv}) } using the method similar to \textbf{Algorithm} \ref{alg:newtonized} are presented in appendix. 
 
 To demonstrate the effectiveness of the proposed method,  \bluetext{ tensors of interest $\m{Y}$ in two scenarios are generated for experiments. In the first case, $L = 2$ sources are identified with $M = 5, N = 4$. In the second case, $L = 16$ sources are identified with $M = 5, N = 32$. In both cases, observations $\mathbf{Y}$ are generated from the following:} 
\begin{equation}\label{eq:signal_model}
    \m{Y} = \sum\limits_{i = 1}^L \m{a}(f_i^1) \otimes \m{a}(f_i^2) \otimes \m{c}_i + \m{N}
\end{equation}
Each entry of $\m{N} \in \mathbb{C}^{N \times N \times M}$ is sampled from circularly symmetric Gaussian distribution $\mathcal{CN}(0, 1)$. All signal tensors \bluetext{ are first sampled from circularly symmetric complex Gaussian distribution and then normalized to  have the same power}, i.e., $\left\|\m{c}_i\right\|_2^2 = P$. The angles $f_i^1, f_i^2$ are independently randomly sampled from uniform distribution over the interval $[0, 2\pi)$. 

\begin{figure}[htbp!]
\centering
\includegraphics[width=.95\linewidth]{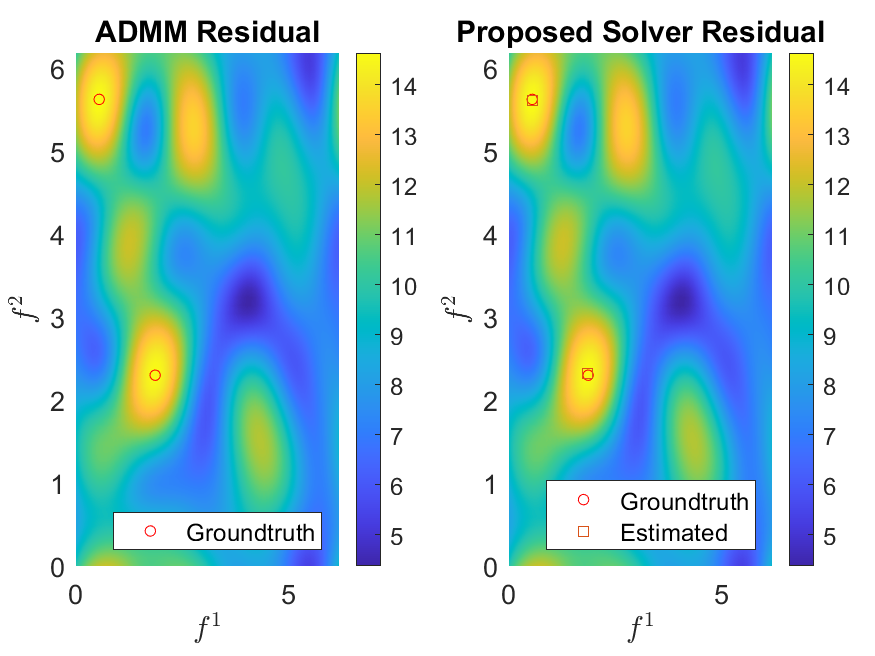}
\caption{\bluetext{Visualization of the spectrum of a residual tensor $\m{Y}_r \in \mathbb{C}^{4 \times 4 \times 5}$ in a random trial with $L = 2$ sources. The solutions returned by the proposed solver and the \ac{ADMM} solver are the same. Note that \ac{ADMM} returns only $\m{X}$. Estimating frequencies $f_i^1, f_i^2$ needs extra calculations for polynomial rooting over $\m{X}$.}}
\label{fig:2D_MMV_residual}
\end{figure}

\bluetext{ The choice of $\zeta$ is related to the noise statistics. In general, based on the thresholding condition in theorem \ref{theorem:anm_v2_optimality}, the rule-of-thumb is always to set $1/\zeta = \mathbb{E}_\m{N} \sup_\m{a \in \mathcal{A}}\left<\m{N}, \m{a}\right> $ \cite{ast}. Notice that it is non-trivial to evaluate: 
\begin{equation}\label{eq:zeta_best_2dmmv}
\mathbb{E}_\m{N} \sup_\m{a \in \mathcal{A}}\left<\m{N}, \m{a}\right> = 
    \mathbb{E}_\m{N}\sup_{f_i^1, f_i^2, \left\|\m{c}_i\right\|_2 = 1}\left<\m{N}, \m{a}(f_i^1) \otimes \m{a}(f_i^2) \otimes \m{c}_i \right>
\end{equation}
Fortunately, since $\m{N}$ is complex random Gaussian matrix and elements in $\mathcal{A}$ all have $\left\|\m{a}\right\|_2^2 = N^2$, the threshold is approximated with mean plus 6 times standard deviation of (\ref{eq:zeta_intermediate}):
\begin{equation}\label{eq:zeta_intermediate}
\begin{split}
    \xi & =  
    \mathbb{E}_\m{N}\sup_{\left\|\m{c}_i\right\|_2 = 1}\left<\m{N}, \m{a}(f_i^1) \otimes \m{a}(f_i^2) \otimes \m{c}_i \right> \leq 
    \mathbb{E}_\m{N} \sup_\m{a \in \mathcal{A}}\left<\m{N}, \m{a}\right>\\
\end{split}
\end{equation}
\begin{equation}\label{eq:zeta_pfa}
\begin{split}
    \frac{1}{\zeta} & = \xi + 6 \sqrt{N^2M - 
    \xi^2}  \\
    & = \frac{N\ \Gamma(M + 1/2)}{\Gamma(M)} + 6\sqrt{N^2M - \frac{N^2\Gamma(M + 1/2)^2}{\Gamma(M)^2}} \\
\end{split}
\end{equation}
}
\bluetext{ In the first case in which $N = 4, M = 5, L = 2$,  the solution returned by the proposed solver is compared to that of solving the \ac{SDP} formulation of two-dimensional \ac{MMV} \ac{AST} \cite{compressive2d} through \ac{ADMM}. For both solvers, a tolerance of $\varepsilon = 10^{-5}$ is allowed. The visualization of the spectrum of the residual tensors returned by the two solvers 
 $\left\|\left<\m{Y}_r, \m{a}(f_1) \otimes \m{a}(f_2)\right>\right\|_2$ are provided in the Figure \ref{fig:2D_MMV_residual}}. As in typical \ac{ANM} problems, the residual tensor $\m{Y}_r$ demonstrates the behavior of a dual {\color{black} certificate of support}. Its spectrum indicates the true frequencies.
 
\begin{figure}[htbp!]
\centering
\includegraphics[width=.95\linewidth]{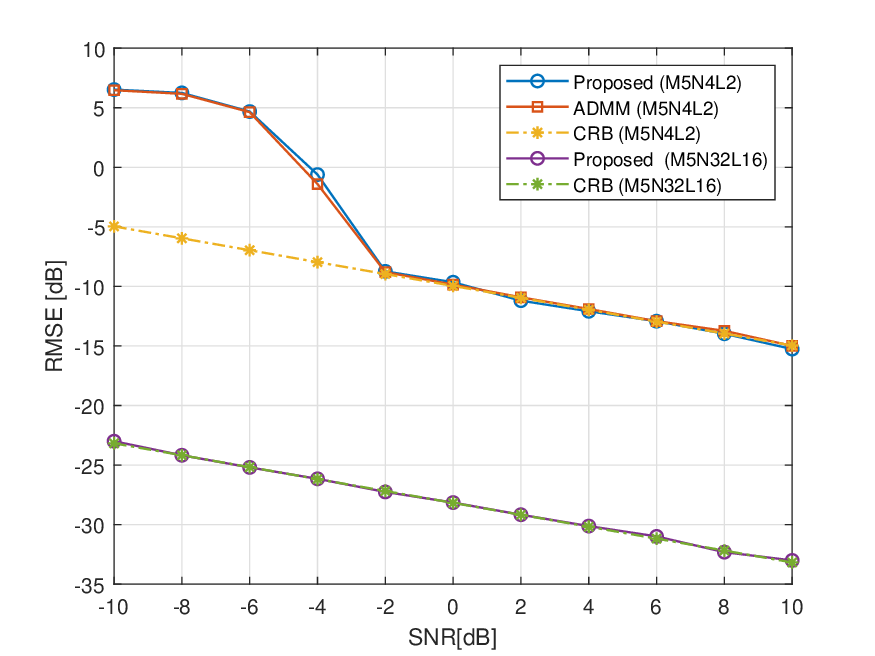}
\caption{Performance of the proposed algorithm on two-dimensional \ac{MMV} \ac{AST}. \bluetext{ Results corresponding to the first case (with $Y \in \mathcal{C}^{4 \times 4 \times 5}$ and $L = 2$ sources) are labeled with $'M5N4L2'$, while the results corresponding to the second case (with $Y \in \mathcal{C}^{32 \times 32 \times 5}$ and $L = 16$ sources) are labeled as $'M5N32L16'$. Each data point is an average of 80 Monte Carlo trials. In each trial, the problem is solved to tolerance $\mathrm{\varepsilon} = 10^{-5}$.} }
\label{fig:2D_MMV_crb}
\end{figure}

\bluetext{ The estimated frequencies $\hat{f}_i^1,$ $\hat{f}_i^2$ are also compared to ground truth at varies \ac{SNR}}. The \ac{SNR} per signal tensor in this experiment is calculated as $\mathrm{SNR} = 20\log_{10}\left(\frac{P}{M}\right)$. 
The error in the estimation process is calculated as:
\begin{equation}\label{eq:doa_err}
    \Delta = \sqrt{\left(\hat{f_i^1} - f_i^1\right)^2 + \left(\hat{f_i^2} - f_i^2\right)^2} 
\end{equation}
which are then compared with the corresponding \ac{CRB} in figure \ref{fig:2D_MMV_crb}. \ac{CRB} in this experiment is calculated from the standard approach in \cite{Trees2002OptimumAP}. The results in figure \ref{fig:2D_MMV_crb} show that $\Delta$ is close to the limit predicted by \ac{CRB} for high enough \ac{SNR} values, which indicates that (\ref{eq:anm_2D_MMV_ANM}) is accurately solved by the proposed solver. \bluetext{ The same experiments are repeated for the second case $M = 5, N = 32, L = 16$, in which the proposed solver can accurately estimate $f^1, f^2$ for all $L = 16$ sources. In the second case the results for \ac{ADMM} solvers are omitted as each iteration involves eigenvalue decomposition of a matrix with  $(M + N^2)^2 = 1029^2$ elements. Thus the overall experiment for \ac{ADMM} solver in the second case is extremely time-consuming.   }

\subsection{Application to weighted \ac{AST}}\label{sec:weighted}

A major weakness of \ac{AST} with \ac{DFT} atomic 
sets (\ref{eq:atomic_set_dft}) is the separation requirement. 
It has been proven that in order for two distinct elements $\m{a}(f_1), \m{a}(f_2) \in \mathbb{C}^N$ to 
be identified in \ac{AST} with high probability, the difference between their frequencies must be large 
enough $\left|f_1 - f_2\right| \geq \frac{4}{N}$ \cite{magazine, candes2014towards, yang2022separation}. To 
overcome such a requirement, a weighted \ac{AST} scheme is proposed in \cite{yang2016enhancing}. Instead of using the vanilla atomic set \bluetext{(\ref{eq:atomic_set_dft}), the following weighted atomic set is
 employed for enhanced the separability of closely spaced frequencies: 
\begin{equation}\label{eq:atomic_set_dft_weighted}
  \left\{w(f)e^{\mathrm{1j}\phi}\m{a}\left(f\right)\left|\phi, f \in \left[0, 2\pi\right), \left[\m{a}\left(f\right)\right]_i = e^{\mathrm{1j}(i - 1)f}\right.\right\}
\end{equation} 
Intuitively speaking, the weighting function $w(f)$ makes some elements from $\mathcal{A}_w$ more favorable than others. Therefore, a carefully designed $w(f)$ can make \ac{AST} a separation-free method of estimating frequencies. In this experiment \cite{yang2016enhancing}, the weighting function was set to $w(f) = \sqrt{\m{a}(f)^H\m{R}^{-1}\m{a}(f)}^{-1}$. The design of the matrix $\m{R}$ is illustrated shortly.}

Let $\mathcal{A}$ be the vanilla set (\ref{eq:atomic_set_dft}) and let $\mathcal{A}_w$ be 
the weighted atomic set (\ref{eq:atomic_set_dft_weighted}). The proposed {\color{black} solver}
 can be applied to {\color{black}\ac{AST}} problems with $\mathcal{A}_w$  as well provided that the first and the    
  second order derivatives $\frac{dw}{df}, \frac{d^2w}{df^2}$ are available.
\bluetext{\begin{align}\label{eq:ast_weighted_smv}
  \underset{L, f_i \in [0, 2\pi), c \in\mathbb{C}}{\mathrm{Minimize}} & \quad 
  \sum\limits_i^L |c_i| + \frac{\zeta}{2} \left\| \m{Y} - \sum\limits_i^L w(f_i)c_i\m{a}(f_i) \right\|_2^2
 \end{align}
}
The key to solve weighted {\color{black}\ac{AST}} problem is \bluetext{ still the following conic projection with shrinkage and thresholding:  
\begin{align}\label{eq:obj_single_atom_weighted}
  \underset{f \in [0, 2\pi) , c\in\mathbb{C}}{\mathrm{Minimize}} & \quad 
  |c| + \frac{\zeta}{2} \left\| \m{Y}_r - w(f)c\m{a}(f) \right\|_2^2
\end{align}
}

\begin{figure*}[htbp!]
   \centering
    \includegraphics[width=.99\linewidth]{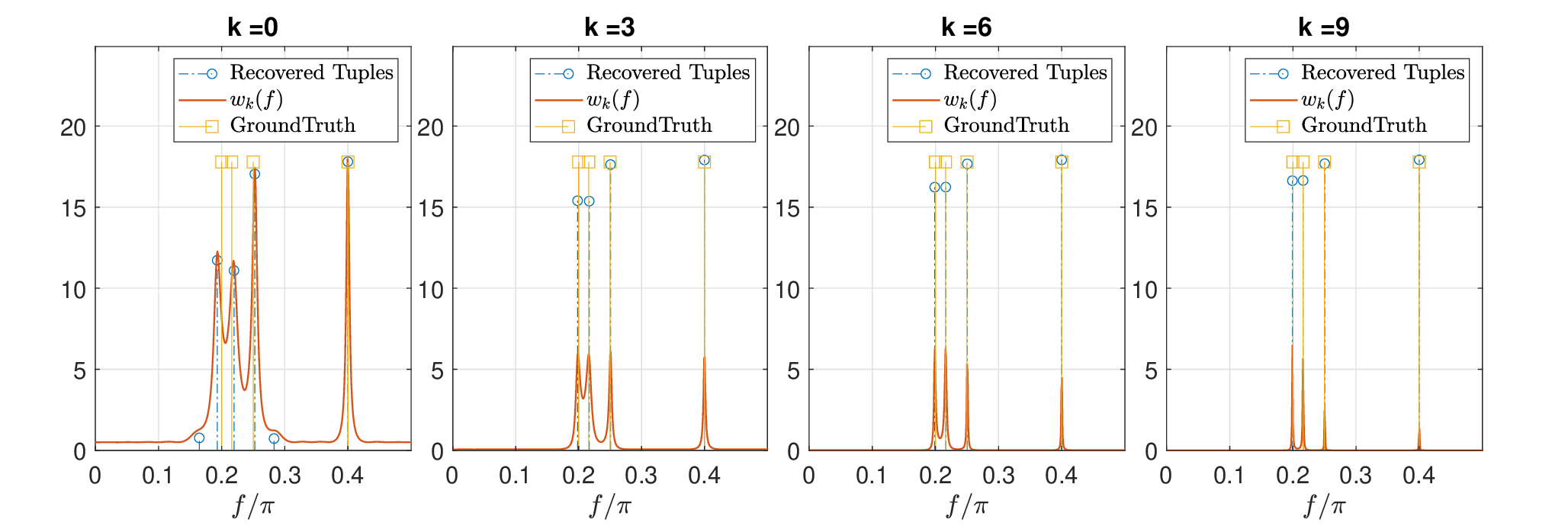}
    \caption{\bluetext{Visualization of the iterative reweighted \ac{AST} process at 25 dB \ac{SNR}. Due to the separation requirement, the outcome of a normal \ac{AST} (0-th iteration) is not accurate.
     With the weighting function based on Capon's 
    beamforming (\ref{eq:anm_w_f}), the closely spaced sources can be successfully differentiated given high enough \ac{SNR}\cite{reweighted2016}}.}
    \label{fig:iterative_reweighted}
\end{figure*}

\begin{figure}[htbp!]
\centering
\includegraphics[width=.95\linewidth]{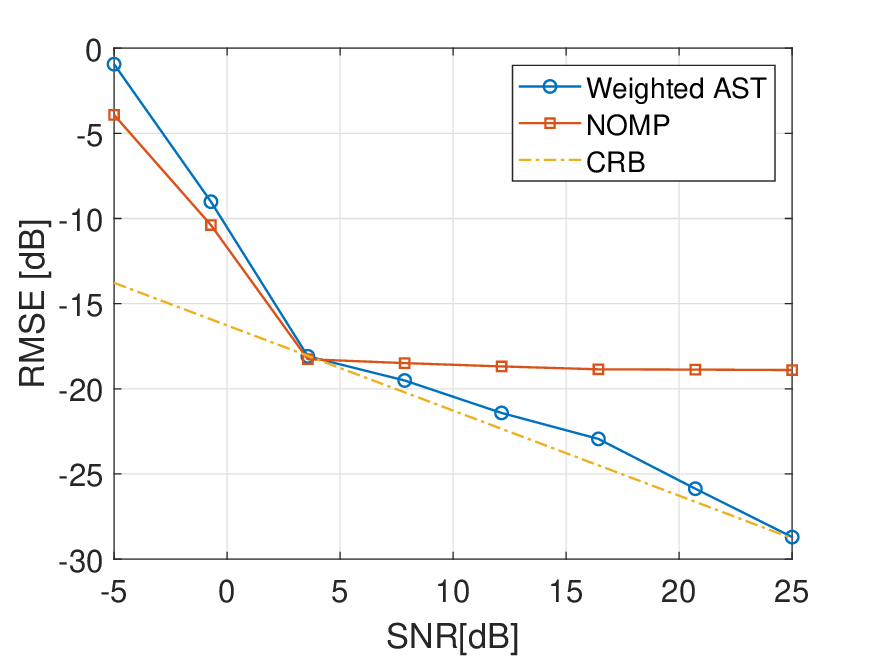}
\caption{\bluetext{Separation-free recovery through iterative reweighted \ac{AST}. In each iteration the weighted \ac{AST} is solved using the proposed solver. For both weighted \ac{AST} and \ac{NOMP}, each data point is an average over 200 Monte Carlo trials.} }
\label{fig:snr_reweighted}
\end{figure}

\bluetext{ Relevant details on obtaining the separable solution (\ref{eq:conic_proj_sol_a}) and (\ref{eq:conic_proj_sol_c}) of (\ref{eq:obj_single_atom_weighted}) are included in appendix \ref{app:weighted_ast}. 
The rest of the subsection presents one of the numerical experiment in 
 \cite{yang2016enhancing}}. The problem is to estimate 
  the ground truth frequencies given a noisy observation: 
\bluetext{
\begin{equation}
    \m{y} = \sum\limits_{i = 1}^L \m{a}(f_i) c_i + \m{n}
\end{equation}
}
Specifically, \bluetext{ $N = 64$ and $L = 5$}. In every Monte Carlo trial each entry of \bluetext{ $\m{n} \in \mathbb{C}^{N}$} is i.i.d. 
standard complex Gaussian random variable. In all trials, $L = 5$ is fixed with 
$\left[f_1, f_2, ..., f_5\right] = 2\pi \left[0.1, 0.108, 0.125, 0.2, 0.5\right]$. To simplify \ac{SNR} calculation, for each source 
\bluetext{ $\left|c_i\right| = P$}.
The $\ac{SNR}$ per source is then defined as $\mathrm{SNR} = 20\log_{10}(P)$. 
Clearly, $f_1, f_2, f_3$ are three closely spaced frequencies that are hard to
differentiate with the $N = 32$ measurement vectors. Based on 
\cite{yang2016enhancing}, the reweighted procedure is used to estimate $f_i$. : 
\begin{itemize}
    \item Starting with $k = 0, \bluetext{\psi_0 = N, \m{R}_0 = \psi_0\m{I}, \zeta_0 = \zeta}$.  
    \item {In the $k$-th iteration, 
    the problem (\ref{eq:ast_weighted_smv}) with \bluetext{$\zeta_k$ and} 
    \begin{equation}\label{eq:anm_w_f}
        w_k(f)  = \left(\m{a}(f)^H\m{R}_k^{-1}\m{a}(f)\right)^{-1/2}
    \end{equation}
     is solved. 
    After obtaining the set $\mathcal{S}_k$ of tuples \bluetext{$(\m{a}(\hat{f_i}), \hat{c}_i)$}, 
    the matrix $\m{R}_{k + 1}$ in  weighting function is updated: 
    \bluetext{\begin{align}
        \psi_{k + 1} & = \psi_k/2 \\
        \m{R}_{k + 1} & = \sum\limits_{i = 1}^{\left|\mathcal{S}_k\right|} \left|\m{c}_i\right| \m{a}(\hat{f_i})\m{a}(\hat{f_i})^\hermitian    
        + \psi_{k + 1}\m{I}
    \end{align}
    The updated weighting function $w_{k+1}(f)$ and $\zeta_{k+1} = \sqrt{2}\zeta_k$ are then used for the next iteration.}
    }
    \item{\bluetext{The process terminates when $\psi_k \leq 10^{-2}$. The solution from the last iteration is returned as estimated results.}}
\end{itemize}
\bluetext{The reweighting process would automatically select detected tuples based on the magnitude of their scalars $|c_i|$. Therefore, different from (\ref{eq:zeta_pfa}), the initial threshold $1/\zeta$ is simply set to the expectation of $\sup_{\left|c\right| = 1}\left<\m{n}, c\m{a}(f)\right>$ without the additional $6$ times of standard deviation: 
\begin{equation}\label{eq:zeta_weighted}
   \frac{1}{\zeta} = \mathbb{E}_\m{n}\sup_{\left|c\right| = 1}\left<\m{n}, c\m{a}(f)\right> = \frac{\sqrt{N\pi} }{2}
\end{equation}
}
\bluetext{The proposed solver is called to solve problem (\ref{eq:ast_weighted_smv}) with $N = 64$ to tolerance $\epsilon = 10^{-3}$ with up to 2000 iterations.} The $0$-th iteration $k = 0$ is the same as 
solving an unweighted {\color{black} AST} problem. From there, the weighting function is updated such that certain elements 
in $\mathcal{A}$ can be better differentiated from others. The result of this reweighted process is demonstrated 
in the figure \ref{fig:iterative_reweighted}. Although the vanilla \ac{AST} fails to differentiate 
the closely spaced frequencies $f_1, f_2, f_3$, the three frequencies are identified 
throughout iterations.   
The overall performance of the weighted \ac{AST} approach for line spectral estimation 
is provided in figure \ref{fig:snr_reweighted}. Each data point is an average of 
20 Monte Carlo trials. 
\bluetext{The performance of the proposed solver for reweighted \ac{AST} problem is also compared to \ac{CRB} and \ac{NOMP}. With the reweighted technique, the accuracy of the estimation 
increases with \ac{SNR}, while the \ac{NOMP} algorithm cannot recover closely spaced frequencies.}  

 \bluetext{The experiment showed that the solver can accurately handle weighted \ac{AST} problems. For reweighted problem with large $N$, the complexity of each reweighted iteration is dominated by evaluating the matrix inversion in $w(f)$. Note that the proposed solver can still be implemented with \ac{FFT} once the matrix $\m{R}$ in $w(f)$ is inverted. This is realized through the Toeplitz structure in $\m{a}(f)^H\m{R}^{-1}\m{a}(f) = \mathrm{trace}(\m{a}(f)\m{a}(f)^H \m{R}^{-1})$.}

\bluetext{
\subsection{Application to Compressive Recovery}

In certain cases \cite{compressive2d}, a full access to the measurement $\m{y}$ is not available. The measurement is observed  through a linear transformation $\m{X}$:
\begin{equation}\label{eq:signal_compressive}
    \m{y} = \sum\limits_{i}^L c_i\m{X} \m{a}(f_i) + \m{n}
\end{equation}
In (\ref{eq:signal_compressive}) $\m{X} \in \mathbb{C}^{M \times N}$ typically has fewer rows than columns $M < N$. In this case, the \ac{AST} problem is formulated as 
\begin{align}\label{eq:ast_compressive}
  \underset{L, f_i \in [0, 2\pi), c \in\mathbb{C}}{\mathrm{Minimize}} & \quad 
  \sum\limits_i^L |c_i| + \frac{\zeta}{2} \left\| \m{y} - \sum\limits_i^L \m{X}c_i\m{a}(f_i) \right\|_2^2
 \end{align}
 The key to solve weighted problem is still the following conic projection with shrinkage and thresholding:  
\begin{align}\label{eq:obj_single_atom_weighted}
  \underset{f \in [0, 2\pi) , c\in\mathbb{C}}{\mathrm{Minimize}} & \quad 
  |c| + \frac{\zeta}{2} \left\| \m{y}_r - \m{X}c\m{a}(f) \right\|_2^2
\end{align}
Using (\ref{eq:conic_proj_sol_a}), (\ref{eq:conic_proj_sol_c}), the projection with shrinkage and thresholding has the following separable structure: 
\begin{align}\label{eq:conic_proj_sol_a_compressive}
    f^* & = \mathrm{argmax}_{f} \quad  \frac{1}{\left\|\m{X}\m{a}(f)\right\|_2}\left(
    \left|\m{y}_r^H \m{X}\m{a}(f)\right| - 1/\zeta\right)\\ \label{eq:conic_proj_sol_a_compressive}
    c^* & = \left\{\begin{matrix}
        0, & \left|\m{y}_r^H \m{X}\m{a}(f)\right| \leq \frac{1}{\zeta} \\[.2em]
      \frac{1 - 1/(\zeta\left|\m{y}_r^H \m{X}\m{a}(f^*)\right|)}{\left\|\m{X}\m{a}(f^*)\right\|_2^2}\m{a}(f^*)^H\m{X}^H\m{y}_r  , & \left|\m{y}_r^H \m{X}\m{a}(f^*)\right| > \frac{1}{\zeta} \\
    \end{matrix}\right.
\end{align}
The effectiveness of the proposed solver  for compressed \ac{AST} problems is verified by experiments on recovering the ground truth frequencies $f_i$ in (\ref{eq:signal_compressive}). Specifically, the solvers are tested with $L = 5, N = 64, M = 30$ and two kinds of compressive matrices $\m{X}$. In the experiments, $c_i$ and $\m{n}$ has the same statistics as that in the previous subsection. The choice of $\zeta$ is set similarly to that in (\ref{eq:zeta_pfa}): 
\begin{equation}
    \frac{1}{\zeta} = \sqrt{M} \left(\frac{\sqrt{\pi} }{2} + 6 \sqrt{1 - \frac{\pi}{4}} \right)
\end{equation}
In the first experiment, $\m{X} \in \mathbb{C}^{M \times N}$ is a random Gaussian matrix, where each entry is sampled from $\mathcal{CN}\left(0, \frac{1}{N}\right)$. In this case $\left\|\m{X}\m{a}(f)\right\|_2$ changes w.r.t $f$ but is largely centered around $\sqrt{M}$. In the second experiment, $\m{X}$ is a selection matrix that randomly picks $M$ out of $N$ entries from $\m{a}(f)$. In this case,  $\left\|\m{X}\m{a}(f)\right\|_2 = \sqrt{M}$ for any $f$. In both experiments the performance of the proposed solver is compared to \ac{CRB} and the \ac{NOMP} algorithm. The results are provided in figure \ref{fig:snr_compressive}. The accuracy of the estimated frequencies verifies that the solver can handle compressive \ac{AST} as well.
\begin{figure}[htbp!]
\centering
\subfloat[Selection Matrix]{\includegraphics[width=.95\linewidth]{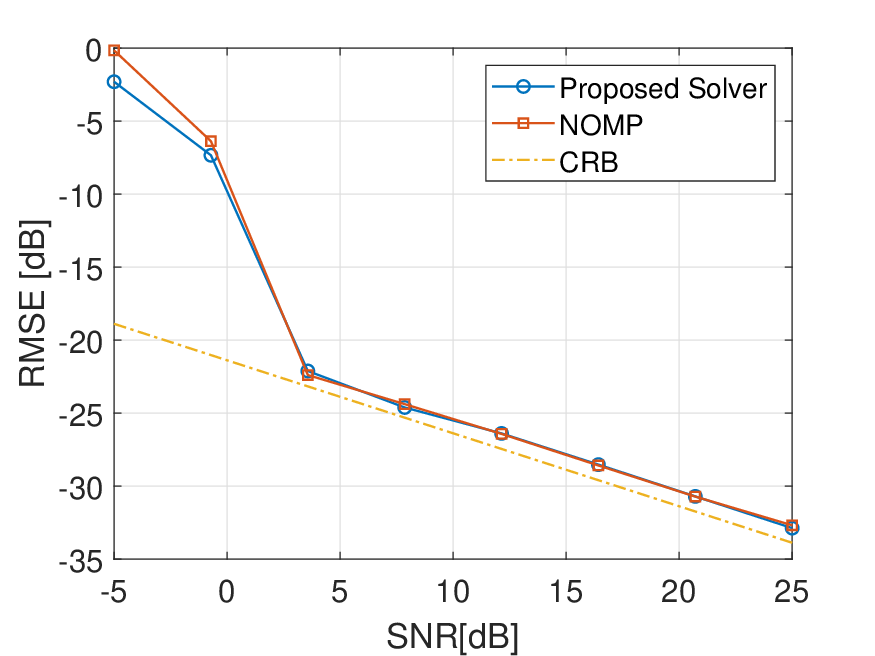}%
\label{fig:snr_compressive_selection}}
\vfil
\subfloat[Gaussian Random Matrix]{\includegraphics[width=.95\linewidth]{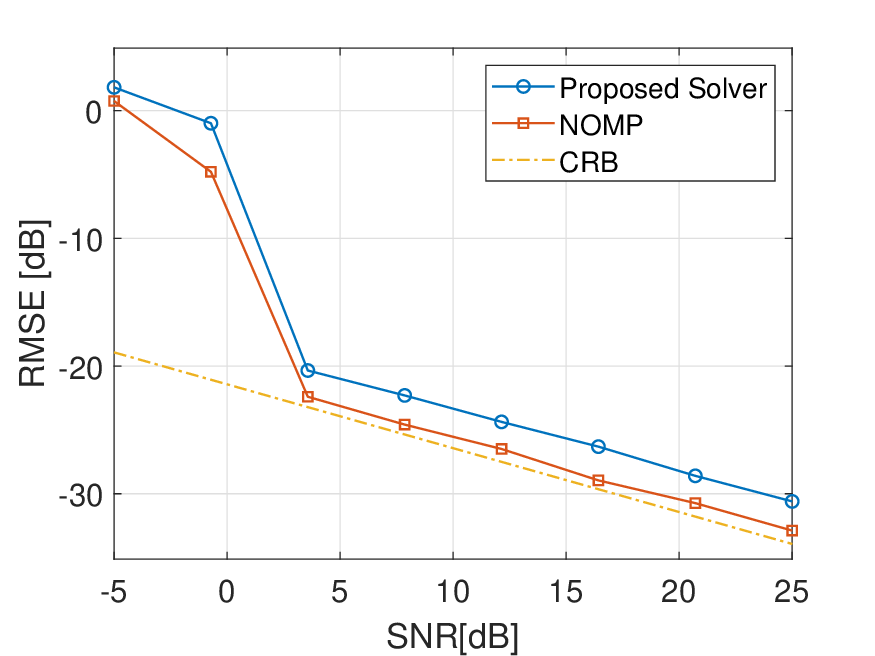}%
\label{fig:snr_compressive_gaussian}}
\vfil 
\caption{\bluetext{Performance of proposed solver in compressive \ac{AST} problems when $N = 64, M = 30, L =5$. In the experiments of the first figure $\m{X}$ selects $M$ out of $N$ elements from $\m{a}(f)$, while in the second $\m{X}$ is a Gaussian random matrix of $M$ rows and $N$ columns.} }
\label{fig:snr_compressive}
\end{figure}
}

\section{Discussion and Future Works}

The major shortcoming of the proposed method is its sensitivity to the sparsity of the problem. The sparsity of {\color{black}\ac{AST} problems are reflected in the minimum number of terms $ L_\mathrm{min}$ need to represent the solutions} and size of the observations $N$. 
In most numerical experiments presented previously, 
the sparsity of the underlying problem is assumed, i.e., $ L_\mathrm{min} \ll N$. {\color{black} This is because algorithm \ref{alg:cyclic} needs at least 
$\mathcal{O}(\left|\mathcal{S}\right|^2)$ iterations to terminate if there are $\left|\mathcal{S}\right|$ 
tuples in $\mathcal{S}$}. Consequently, for a sparse {\color{black}\ac{AST} problem with $L$ terms and} the simplest atomic set (\ref{eq:atomic_set_dft}), the proposed 
coordinate descent {\color{black} solver} yields computational complexity $\mathcal{O}\left(L^2N\log N\right)$. Thus, 
for non-sparse {\color{black} \ac{AST}} problem the coordinate descent 
method becomes highly inefficient. A typical example is the trials in section \ref{sec:tt} with $\zeta = 1/\sqrt{N}$ 
instead of $\zeta = 1/\sqrt{N\log N/4}$. The choice $\zeta = 1/\sqrt{N}$ yields $L_{\mathrm{min}} \sim \mathcal{O}
\left(N^2\right)$, which makes the overall complexity for the proposed method $\mathcal{O}\left(N^3\log N\right)$. 
In such cases, conventional interior-point solvers or \ac{ADMM} for the \ac{SDP} formulation of are still preferred 
as they are not sensitive to the sparsity of the problems. 

Part of the advantage of the coordinate descent {\color{black} solver} is its flexibility. In practice, there are many atomic 
sets with which the corresponding \ac{AST} problems cannot be easily formulated as \ac{SDP}. In such cases, the 
conventional disciplined interior-point solvers are not available. These problems can still be solved 
by the proposed coordinate descent framework as long as sub-problems (\ref{eq:new_tuple_descent}) {\color{black} or (\ref{eq:tuple_coor_descent})} are solvable. Under such assumptions, more {\color{black} \ac{AST}} 
problems become traceable. A recently developed example is \cite{McRae2023optimal} where the authors employed 
a carefully developed atomic set for sparse phase retrieval and principal component analysis. In \cite{McRae2023optimal}, 
a generic method similar to the proposed coordinate descent framework was used simply because that the \ac{SDP} formulation 
is intractable. Although the sub-problem analogous to (\ref{eq:new_tuple_descent}) was also too difficult 
to solve such that the authors only solved it approximately, the performance of their generic method was still better than most 
presented baselines for sparse phase retrieval. 

There are plenty of open research questions along the track of using coordinate descent method to solve \ac{ANM} problems. From the 
perspective of applications, 
many previously untraceable atomic sets can now be considered. For instance, the atomic set of interest 
can be a continuous manifold produced by a pre-trained neural network for a specific image processing 
task. From the perspective of optimization, the proposed coordinate descent solver can be accelerated with 
better initialization. The method itself can also be further optimized to get rid of the ambiguity mentioned 
in remark \ref{remark:ambiguity} or to work with a lower oversampling ratio $r = 8$ with the help of a modified 
algorithm \ref{alg:newtonized} that uses a line-search strategy, etc.   

\section{Conclusion} 

In this paper, an efficient {\color{black} solver}
for sparse atomic norm \bluetext{denoising} is proposed. The method is based on 
the classic idea of coordinate descent. It bypasses the \ac{SDP} formulation of {\color{black} \ac{AST}} 
in conventional convex optimization and converges to the global optimal point by solving 
non-convex \bluetext{projections onto a conic set}. The method {\color{black} has low complexity} and thus offers 
a scalable solution for large-scale {\color{black} \ac{AST}} problems. Several numerical experiments are presented 
to confirm the flexibility and efficiency of the coordinate descent {\color{black} solver}. To conclude, the {\color{black} solver} 
can be a potential candidate for a wide range of applications of \bluetext{super-resolution estimation and denoising}. 

\section{Acknowledgement}

\bluetext{The authors would like to thank the anonymous reviewers for their valuable comments on the technical details, organization, and experiments of the manuscript.}

\bibliographystyle{IEEEbib}
\bibliography{mybib}

\appendices

\section{Solving Two-dimensional \ac{MMV} {\color{black}\ac{AST}}}

\bluetext{ The key to solve two-dimensional \ac{MMV}} {\color{black}\ac{AST}} problem are 
\bluetext{ conic projection with shrinkage and thresholding (\ref{eq:tuple_coor_descent}), (\ref{eq:new_tuple_descent}). The two problems have the same formulation as following:} 
\begin{align}\label{eq:simpified_single_atom_2d_mmv}
    \underset{\m{c}_i \in \mathbb{C}^{M}, f_i^1, f_i^2}{\mathrm{Minimize}} & \quad 
    \left\|\m{c}_i\right\|_2 -\zeta\left<\m{Y}_r,\m{a}(f_i^1) \otimes \m{a}(f_i^2) 
    \otimes \m{c}_i \right> + \frac{\zeta N^2\left\|\m{c}_i\right\|_2^2}{2}
\end{align}
The inner-product in the vector space of three-dimensional tensor is defined as 
\begin{equation}\label{eq:tensor_inner_prod}
    \left<\m{X},\m{Y}\right> = \sum\limits_{i,j,k} \mathrm{Re}\left\{\left[\m{X}\right]_{i,j,k}^*\left[\m{Y}\right]_{i,j,k}
    \right\}    
\end{equation}
Let $\hat{\m{c}}  = \m{c}/\left\|\m{c}\right\|_2$. As in (\ref{eq:opt_f})-(\ref{eq:opt_phi}),
 (\ref{eq:simpified_single_atom_2d_mmv}) also admits 
a separable solution: 
\begin{align}\label{eq:opt_f_2d_mmv}
    (f_1^*, f_2^*) & = \mathrm{argmax}_{f_1, f_2} \left\|\m{Y}_r \odot (\m{a}(f_1) \otimes \m{a}(f_2) \otimes \m{1})\right\|_2 \\ \label{eq:opt_hatc}
    \hat{\m{c}}^* & = \underset{\hat{\m{c}}, \left\|\hat{\m{c}}\right\|_2 = 1}{\mathrm{argmax}} 
    \left\|\m{Y}_r \odot (\m{a}(f_1^*) \otimes \m{a}(f_2^*) \otimes \hat{\m{c}})\right\|_2
\end{align}
The role of $\left\|\m{c}\right\|_2$ is the same as that of $c$ in (\ref{eq:opt_c}). The optimal 
$\left\|\m{c}\right\|_2^*$ involves a shrink and thresholding step. Let $\eta = 
\left\|\m{Y}_r \odot (\m{a}(f_1^*) \otimes \m{a}(f_2^*) \otimes \hat{\m{c}}^*)\right\|_2$. Then, 
\begin{equation}\label{eq:opt_c_2d_mmv}
\left\|\m{c}\right\|_2^* = \left\{\begin{matrix}
    0, & \ \eta \leq \frac{1}{\zeta} \\
  \frac{1}{N^2}\left(\eta - \frac{1}{\zeta}\right), & \ \eta > \frac{1}{\zeta} \\
\end{matrix}\right.    
\end{equation}
The key step is still (\ref{eq:opt_f_2d_mmv}). Similar to algorithm \ref{alg:newtonized}, the 
following procedure is employed to solve (\ref{eq:opt_f_2d_mmv}): 
\begin{itemize}
    \item{To start with, an oversampled two-dimensional \ac{FFT} is performed on the first two dimensions 
    of $\m{Y}_r$ to evaluate 
    $\left\|\m{Y}_r \odot (\m{a}(f_1) \otimes \m{a}(f_2) \otimes \m{1})\right\|_2$ on a fine mesh of grid 
    points $(f_1, f_2)$. The coordinate of the maximum over the mesh is used as the starting point.}
    \item{Starting with the maximum $(f_1, f_2)$ over the mesh, use Newton's method to solve for 
    the off-grid maximum until convergence.}
\end{itemize}
\bluetext{ With an over-sampled initial search}, the Newton's method can converge 
to the global maximum $(f_1^*, f_2^*)$ of the non-convex function $\left\|\m{Y}_r \odot (\m{a}(f_1) \otimes \m{a}(f_2) \otimes \m{1})\right\|_2$
Plugging in $(f_1^*, f_2^*)$ to (\ref{eq:opt_hatc}), (\ref{eq:opt_c_2d_mmv}) then completes the solution to 
(\ref{eq:simpified_single_atom_2d_mmv}).

\section{Solving Weighted {\color{black}\ac{AST}}}\label{app:weighted_ast}

The discussion here generalizes the case presented in \ref{sec:weighted} to \ac{MMV}, i.e., $M \geq 1$ and $c \in \mathbb{C}$ is replaced by $\m{c} \in \mathbb{C}^M$. The key to solve weighted \ac{MMV} {\color{black}\ac{AST}} problem is 
the conic projection (\ref{eq:obj_single_atom_weighted}), which is then simplified as: 
\begin{align}\label{eq:simplified_single_atom_weighted}
    \underset{\m{c} \in \mathbb{C}^{M}, f}{\mathrm{Minimize}} & \quad 
    \left\|\m{c}\right\|_2 -\zeta\left<\m{Y}_r,w(f)\m{a}(f) 
    \otimes \m{c} \right> + \frac{\zeta w^2(f){\color{black} N}\left\|\m{c}\right\|_2^2}{2}
\end{align}
Different from unweighted cases of \ac{DFT} atomic set, the second order term in (\ref{eq:simplified_single_atom_weighted}) also depends on $f$ because of the weights $w^2(f)$. Let $\hat{\m{c}} = 
\m{c}/\left\|\m{c}\right\|_2$. The objective function in  (\ref{eq:simplified_single_atom_weighted}) can be re-organized as: 
\begin{align}\label{eq:reorganized_single_atom_weighted}
     & \left\|\m{c}\right\|_2 - \zeta \left<\m{Y}_r, w(f)\m{a}(f)
    \otimes \m{c}\right> + \frac{\zeta w^2(f){\color{black} N}\left\|\m{c}\right\|_2^2}{2} \\ \nonumber 
    = & w(f)\zeta\left\|\m{c}\right\|_2 \left(\frac{1}{\zeta w(f)} -  
    \left<\m{Y}_r, \m{a}(f)
    \otimes \hat{\m{c}}\right>\right) 
    + \frac{\zeta w^2(f){\color{black} N}\left\|\m{c}\right\|_2^2}{2}\\ \nonumber
\end{align}
(\ref{eq:simplified_single_atom_weighted}) indicates a separable solution for (\ref{eq:obj_single_atom_weighted}) as following: 
\begin{align}\label{eq:opt_f_weighted_mmv}
    f^*& = \mathrm{argmax}_{f} \left\|\m{Y}_r^H \m{a}(f)\right\|_2 - \frac{1}{\zeta w(f)} \\ \label{eq:opt_hatc_weighted}
    \hat{\m{c}}^* & = \underset{\hat{\m{c}}, \left\|\hat{\m{c}}\right\|_2 = 1}{\mathrm{argmax}}  
    \left<\m{Y}_r, \m{a}(f^*) \otimes \hat{\m{c}}\right> 
\end{align}
The shrinkage and thresholding step on $\left\|\m{c}\right\|_2$ becomes slightly different as it involves the weighting function $w(f)$. Let $\eta = \left<\m{Y}_r, \m{a}(f^*) \otimes \hat{\m{c}}^*\right>$. The optimal $\left\|\m{c}\right\|_2^2$ is calculated from the following: 
\begin{equation}\label{eq:opt_c_weighted_mmv}
\left\|\m{c}\right\|_2^* = \left\{\begin{matrix}
    0, & \ \eta \leq \frac{1}{w(f)\zeta} \\
  \frac{1}{w(f){\color{black} N}}\left(\eta - \frac{1}{w(f)\zeta}\right), & \ \eta > \frac{1}{w(f)\zeta} \\
\end{matrix}\right.    
\end{equation}
(\ref{eq:opt_c_weighted_mmv}) clearly indicates the functionality of $w(f)$ which has an impact on the threshold $1/\zeta$. Unfortunately, $w(f)$ also makes it harder to solve the key step (\ref{eq:opt_f_weighted_mmv}) as \ac{FFT} along is not enough to evaluate $\left\|\m{Y}_r^H \m{a}(f)\right\|_2 - \frac{1}{\zeta w(f)}$ on a large number of points. However, the methodology remains the same: 
\begin{itemize}
    \item{To start with, an oversampled two-dimensional \ac{FFT} is performed on the first dimension 
    of $\m{Y}_r$ to evaluate 
    $\left\|\m{Y}_r^H \m{a}(f) \right\|_2$ on a fine grid points over $[0, 2\pi)$. Additionally, $\frac{1}{w(f)\zeta}$  is also evaluated over the same fine grid.     
    The coordinate of the maximum $\left\|\m{Y}_r^H \m{a}(f)\right\|_2 - \frac{1}{\zeta w(f)}$ over the grid is used as the starting point.}
    \item{Starting with the maximum on-grid $f$, use Newton's method to solve for 
    the off-grid maximum until convergence. In each Newton's step, the first and the second order derivatives $\frac{dw}{df}$, $\frac{d^2w}{df^2}$ need to be computed.}
\end{itemize}
Once the off-grid maximum $f^*$ is found, plugging $f^*$ into (\ref{eq:opt_c_weighted_mmv}), (\ref{eq:opt_hatc_weighted}) completes the solution.

\end{document}